\documentclass[reprint,aps,pre]{revtex4}
\usepackage[a4paper, margin=1in]{geometry}
\usepackage{graphicx}
\usepackage{natbib}
\usepackage{color}
\usepackage{amsmath}
\usepackage{ulem}
\usepackage[bookmarks=false]{hyperref}

\begin{document}

\title{Modelling droplet--particle interactions on solid surfaces by coupling the lattice Boltzmann and discrete element methods}

\author{Abhinav Naga}
\email{abhinav.naga@ed.ac.uk}
\author{Xitong Zhang}
\author{Junyu Yang}
\author{Halim Kusumaatmaja}
 \email{halim.kusumaatmaja@ed.ac.uk}

\affiliation{Institute for Multiscale Thermofluids, School of Engineering, University of Edinburgh, Edinburgh, EH9 3FD, United Kingdom}

\begin{abstract}

We develop a three-dimensional numerical scheme for investigating interfacial flows coupled with frictional solid particles. Our approach combines the lattice Boltzmann method (LBM) to model the dynamics of a two-component fluid, and the discrete element method (DEM) to model normal reaction, sliding friction, and rolling friction between solid particles and between particles and solid surfaces. Key to the coupling between the fluid and particle dynamics are the momentum exchange method to transfer hydrodynamic forces between the fluids and particles, a geometric boundary condition to tune particle wettability, and a capillary force model describing surface tension forces between particles and liquid–fluid interfaces. We rigorously validate the contact forces by investigating the dynamics of a particle bouncing off a solid surface and rolling down an inclined plane, the hydrodynamic force by the Segr\`{e}-Silberberg effect, and the capillary force by particle detachment from a liquid-fluid interface. Motivated by the self-cleaning properties of lotus leaves, we apply the method to investigate how drops remove contaminant particles from surfaces. We successfully reproduce scenarios reported experimentally by Naga \textit{et al.} (\textit{Soft Matter} (2021) 17(7):1746-1755) by tuning the particle friction. Furthermore, the LBM-DEM approach allows us to systematically explore the effects of particle friction coefficients, drop size, and speed. Our method opens opportunities to study numerous phenomena involving particle dynamics interacting with interfacial flows, including soil erosion, capillary-driven colloidal self-assembly, and how raindrops transport microplastics in the environment. It also makes it possible to control parameters that are difficult to tune independently in experiments, including contact angles, surface tension, and friction coefficients.
\end{abstract}
\maketitle

\section{Introduction}

Multiphase flows consisting of two immiscible fluids and solid particles control many natural and industrial processes. Some examples include the process of rain or fog droplets removing dirt particles from self-cleaning lotus leaves \citep{barthlott_purity_1997, furstner_wetting_2005}, the removal of microplastics from wastewater by flotation \citep{monira_nano_2023, jia_advanced_2024}, the process of soil erosion when raindrops impact the ground \citep{vaezi_contribution_2017}, and the safe storage of carbon dioxide under the seabed \citep{yeilding_deepwater_2022}. The physics underpinning all these problems involves the interactions between one or more fluids and solid particles. In these processes, several forces are involved, including hydrodynamic forces, capillary forces between particles and liquid--fluid interfaces, and friction between particles. These interactions are often difficult to decouple experimentally. For example, changing the surface chemistry of particles typically influences both the friction force between two particles as well as their wettability towards liquids. Changing the particle roughness influences both the particle wettability \citep{dettre_contact_1964} and its ability to slide or roll on a solid surface \citep{singh_shear_2020}. Numerical simulations are valuable to decouple these interactions and study their influence independently. In this paper, we propose a numerical method to investigate problems involving up to two immiscible fluids with rigid frictional solid particles and solid substrates. The method allows us to independently tune fluid viscosities, surface tension, sliding and rolling friction, and wettability. We show that the method is a promising tool for investigating problems in which hydrodynamic, capillary, and friction forces all have important contributions.

To explicitly model two-component fluid flows coupled with solid particles, we must account for several interactions. We need to solve for the fluid flow and the interfacial dynamics between the fluids, couple hydrodynamic force from the fluid flow to the solid particle, account for capillary forces when the particle is at the interface, and incorporate normal reaction and friction forces between solid surfaces that come in contact. When particles can roll, rolling friction must also be considered. For several problems, two-dimensional (2-D) models are insufficient and these interactions must be modelled in three dimensions. For example, when drops remove particles from self-cleaning surfaces, the particle may move from the front to the rear of the drop, following the circular drop footprint without fully entering the drop \citep{naga_how_2021}. This behaviour cannot be obtained in two dimensions, because in two dimensions the only way for the particle to reach the rear side is by fully entering the drop at the front. Another example where it is important to use three-dimensional (3-D) models is the study of the process of wet granulation, used by industries such as the pharmaceutical and food industries \citep{singh_challenges_2022}. Wet granulation involves using drops to aggregate fine powders into larger granules. It is important to consider 3-D geometries in this process because the packing fraction is larger in two dimensions than in three dimensions \citep{eales_properties_2023}.

Solid-solid interactions between particles and solid substrates are typically modelled using the discrete element method (DEM) \citep{cundall_discrete_1979} or by applying repulsive potentials when the separation between two solids becomes less than a prescribed threshold distance. Traditionally, DEM has been applied to study the behaviour of frictional and cohesive powders or granular media. The DEM involves solving Newton's second law for the translational and rotational motion of every particle, given the contact forces and torques acting on the particle. Compared with continuum methods (\textit{e.g.} finite element), DEM can provide detailed insights into the dynamics of representative volume elements of granular media at a single-particle level. Since the first realistic model for sliding friction was proposed by \citet{cundall_discrete_1979}, several extensions and alternatives have been proposed for inter-particle adhesion forces, sliding friction, rolling friction, and torsion resistance. These models vary in their level of complexity. In this work, we choose a contact force model that provides a good compromise between a realistic and an easy-to-handle approach~\citep{luding_cohesive_2008}.  In this model, whenever two particles come into contact, a virtual linear spring-dashpot emerges in the contact region and resists relative motion in the normal and tangential directions. Rolling friction is modelled similarly to restrain rolling motion that can in practice arise from surface roughness, surface deformation, and asymmetries in the shape of the contact when particles roll.

Methods to model fluid-fluid interactions and account for the interface between the fluids include continuum methods (\textit{e.g.} volume of fluid and level set method), mesoscale methods (\textit{e.g.} the lattice Boltzmann method (LBM)) and molecular-level methods (\textit{e.g.} molecular dynamics). The LBM has emerged as a powerful tool for simulating complex interfacial flows in the past three decades. The LBM is a mesoscopic method based on kinetic theory. The basic idea behind the LBM is that the fluid is modelled as a collection of fictive fluid elements, described by distribution functions \citep{kruger_lattice_2016, succi_lattice_2018}. These fluid elements lie at discrete positions on a lattice mesh. In each time step, these fluid elements propagate along one of several possible discrete velocity directions and collide with other fluid elements according to a set of carefully derived rules such that mass, momentum, and energy are conserved. It can be shown that the LBM approach is equivalent to solving the Navier-Stokes equations \citep{kruger_lattice_2016}. Most operations are local in LBM, making it highly parallelizable and compatible with DEM. Although LBM is now an established method to study interfacial flows and wetting dynamics on static solid surfaces with complex solid geometries \citep{kusumaatmaja_moving_2016,wohrwag_ternary_2018} and frictionless colloidal particles \citep{stratford_colloidal_2005,stratford_lattice_2005,cates_simulating_2004,cates_physical_2005}, very few methods have attempted to couple multiphase LBM with frictional solid particles \citep{fei_coupled_2023, jiang_coupled_2022}.

In general, algorithms that couple solid-solid interactions, fluid-fluid interactions, and solid-fluid interactions are complex and computationally demanding, especially in three dimensions. Consequently, most existing algorithms for coupling these interactions are limited to single-phase flows and/or 2D geometries \citep{boutt_coupled_2011,peng_influence_2014,kafui_discrete_2002,li_numerical_2021,yang_comprehensive_2019,lomine_modeling_2013,cui_coupled_2014,owen_efficient_2011,mansouri_numerical_2017}. A limited number of algorithms have been proposed to couple DEM and two-fluid flows \citep{washino_development_2023,davydzenka_coupled_2020,xia_computational_2024}. However, these algorithms are typically restricted to 2D geometries \citep{fei_coupled_2023,jiang_coupled_2022}, and do not explicitly account for capillary forces \citep{shen_resolved_2022,pozzetti_multiscale_2018,lai_signed_2023,kano_numerical_2020,van_sint_annaland_numerical_2005,chu_cfd-dem_2009} when particles are at interfaces or rolling friction when particles roll against one another \citep{ge_3-d_2006,baltussen_direct_2013}.

In this paper, we propose a coupled LBM--DEM method that accounts for all these forces in three dimensions. The key novelty is that our method explicitly accounts for hydrodynamic, capillary, and friction forces, allowing us to tune the liquid viscosity, surface tension, particle wettability, and sliding and rolling friction coefficients independently. We rigorously benchmark all the forces involved. To benchmark the normal reaction and friction forces, we analyse the motion of a solid particle bouncing and moving on a flat solid surface. For the hydrodynamic force, we compare the motion of a particle in a Poiseuille flow with experimental results demonstrating the Segrè–Silberberg effect. For the capillary force, we measure the force required to detach a particle from a liquid-fluid interface and show that our results are in good agreement with analytical predictions.

Our method opens the possibility to simulate how drops remove particles from solid surfaces, providing insights into how to design self-cleaning surfaces. In particular, the method is ideally suited to study this problem because during particle removal, friction and capillary forces can have the same order of magnitude and thus none of these forces can be neglected~\citep{naga_how_2021}. Our numerical results are in good agreement with previous experiments that have imaged and quantified the horizontal force between a water drop and a hydrophobic particle on a hydrophobic surface~\citep{naga_how_2021}. Our results reveal that when a drop removes a particle, the particle can either slide or roll, depending on the relative magnitude of the coefficients of sliding and rolling friction. Our systematic parameter variation also indicates that a successful removal depends on several factors, including particle friction, drop size, and drop speed.


\section{Overview of macroscopic equations of motion for the fluids and particles}

In this section, we give an overview of the equations that describe the macroscopic behaviour of the two fluids and of the solid particles. In the next sections, we provide further details on the algorithms that we use to solve for the fluid dynamics, particle dynamics, and the coupling between the fluids and the particle.

For simplicity, here we consider two incompressible fluids with equal density and tuneable viscosity ratio. Throughout this paper, we focus on problems where viscous and capillary forces dominate inertial forces. Since inertial forces are not significant in these problems, the equal fluid density model is sufficient to capture the dominant interactions. We will refer to the two fluids as `liquid' (fluid \textit{a} with high viscosity) and `air' (fluid \textit{b} with low viscosity). Note that the density of the particle is independent of the fluid density and can be tuned freely to vary the mass and moment of inertia of the particle.

In a continuum framework, the governing equations for the velocity and pressure fields for the two fluids are the Navier-Stokes equations,
\begin{equation}
\nabla\cdot\boldsymbol{u}=0, \label{NS1}
\end{equation}
\begin{equation}
\rho_f\left[\frac{\partial \boldsymbol{u}}{\partial t} + (\boldsymbol{u\cdot\nabla)\boldsymbol{u}}\right]=-\nabla p+\nabla\cdot[\eta(\phi)(\nabla\boldsymbol{u}+(\nabla\boldsymbol{u})^T )]+\boldsymbol{f}_\sigma(\phi)+\boldsymbol{f}_g(\phi).
\label{NS2}
\end{equation}
Here, $\boldsymbol{u}$, $\rho_f$, $p$, and $\eta$, are the local fluid velocity, density, pressure, and dynamic viscosity, respectively. The value of the fluid viscosity at a given point depends on the order parameter, $-1\leq \phi \leq 1$, that characterises the distribution of the two fluids within the domain. The last two terms, $\boldsymbol{f}_\sigma$ and $\boldsymbol{f}_g$, correspond to forces (per unit volume) arising from surface tension and gravity, respectively. The surface tension force $\boldsymbol{f}_\sigma$ depends on the order parameter and is only active where the two fluids meet. The gravitational force $\boldsymbol{f}_g$ also depends on the order parameter to give the option of applying a body force to one of the fluids only. Therefore, although the two fluids have equal density, it is still possible to mimic the effect of gravity (or buoyancy) between the fluids by applying a body force to one of the fluids only. The order parameter evolves in tandem with the fluid velocity according to the advection equation,
\begin{equation}
    \frac{\partial \phi}{\partial t}+\boldsymbol{u}\cdot\nabla \phi=0.
\end{equation}
The order parameter allows us to identify how the two fluids are distributed within the domain and locate the position of interfaces between them. This information is used to compute the surface tension force $\boldsymbol{f}_\sigma$ as well as the local fluid viscosity. 

The equation for the translational motion of a particle that is subject to a net force is given by Newton's second law,
\begin{equation}
  m_p\frac{\mathrm{d}^2\boldsymbol{x}_p}{\mathrm{d}t^2}=\boldsymbol{F}_p=\boldsymbol{F}_g+\boldsymbol{F}_\eta +\boldsymbol{F}_c+\boldsymbol{F}_\sigma.
  \label{Newton second law}
\end{equation}
Here, $m_p$ is the mass of the particle, and $\boldsymbol{x}_p$ is its position. The terms on the right-hand side denote the gravitational force $\boldsymbol{F}_g$, the hydrodynamic force $\boldsymbol{F}_\eta$ due to momentum transfer between the particle and the surrounding fluid, the contact forces $\boldsymbol{F}_c$ (normal reaction and friction) when a particle makes contact with a solid particle or substrate, and the capillary force $\boldsymbol{F}_\sigma$ due to the direct action of surface tension at solid-liquid-air three-phase contact line when the particle is at the interface between the two fluids. The contact forces include contributions from all contact points, $\boldsymbol{F}_c=\sum_{i}\boldsymbol{F}_c^i$, where $\boldsymbol{F}_c^i$ is the force exerted on the particle by the contacting solid $i$, which can be another particle or a solid substrate, and the summation runs over all contacts. 

The equation for the rotational motion is given by
\begin{equation}
  I_p\frac{\mathrm{d}^2\boldsymbol{\varphi}_p}{\mathrm{d}t^2}=\boldsymbol{T}_p=\boldsymbol{T}_\eta+\boldsymbol{T}_c +\boldsymbol{T}_\sigma,
  \label{Newton - rotation}
\end{equation}
where $I_p$ is the moment of inertia of the particle, $\boldsymbol{\varphi}_p$ is the angular displacement. The terms on the right-hand side, $\boldsymbol{T}_\eta$, $\boldsymbol{T}_c$, and $\boldsymbol{T}_\sigma$, denote the torque due to hydrodynamic force, contact forces, and capillary force, respectively. The torque due to contact forces includes contributions from sliding friction and rolling friction. Rolling friction affects the rotational motion only and not the sliding motion, thus, it is only included in the equation for rotational motion. Since gravity acts uniformly throughout the particle, it does not generate a torque on the particle.

We use the Verlet integration algorithm to solve Newton's second law (equations \ref{Newton second law}, \ref{Newton - rotation}). With this algorithm, energy is conserved during collisions by solving Newton's laws based on an average of the forces/torques in the previous and current time step.

The interaction force that the particle exerts on the fluid enters the Navier-Stokes equations as boundary conditions. We use the no-slip and no-penetration boundary conditions at solid walls. Additionally, when the particle is at the liquid-air interface, we impose a wetting boundary condition that sets the equilibrium contact angle between the particle and the liquid. This contact angle enters the calculation of the capillary force and capillary torque, $\boldsymbol{F}_\sigma$ and $\boldsymbol{T}_\sigma$.

\section {Discrete element method for modelling contact forces}

In this section, we describe how we model the contact forces $\boldsymbol{F}_c$, and the contact torques $\boldsymbol{T}_c$ when two solid surfaces make contact.  There are two types of contact forces: normal reaction and friction. Friction includes both sliding and rolling friction. We explicitly account for two forms of contact: contact between a spherical particle and a flat solid substrate, and contact between two spherical solid particles.

In DEM, solids are rigid but their contacts are `soft', which means that they are allowed to overlap by a small amount. They only exert contact forces on one another when they overlap. The purpose of the DEM model is to compute the contact force $\boldsymbol{F}_c$, and contact torque $\boldsymbol{T}_c$ that enter Newton's laws for the translational and rotational motion of the particles (equations \ref{Newton second law} and \ref{Newton - rotation}). 

\subsection{Normal reaction}

\begin{figure}
  \centerline{\includegraphics[width=1.0\linewidth]{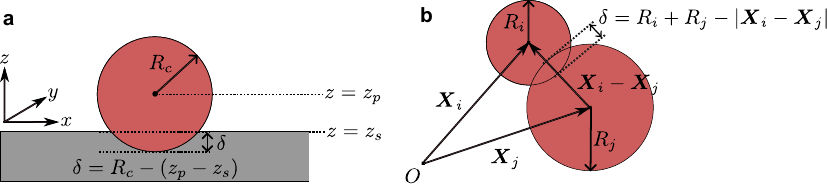}}
  \caption{Definition of the overlap between two solids in contact. (a) Definition of the overlap between a spherical particle and a flat substrate. (b) Definition of the overlap between two spherical particles.}
\label{fig:OverlapDefinition}
\end{figure}

When a particle comes into contact with a flat solid surface, we define the overlap as (see figure \ref{fig:OverlapDefinition}a)
\begin{equation}
\delta = R_c -(z_p-z_s),
\label{Overlap}
\end{equation}
where $R_c$ is the mechanical radius of the particle, $z_p$ is the $z-$coordinate of the centre of the particle and $z_s$ is the $z-$coordinate corresponding to the top of the solid substrate. The contact force is modelled with a spring-dashpot model. For a perfectly elastic contact without any dissipation, the normal contact force is given by $F^0_n=k_n\delta$, where $k_n$ is the spring stiffness for overlap in the normal direction. But since energy is generally dissipated during realistic collisions, a damping term is added to the expression such that the magnitude of the normal reaction force becomes
\begin{equation}
F_n=k_n\delta+ \gamma_nv_n.
\label{Contact force}
\end{equation}
Here, the second term is a damping term, where $\gamma_n$ is the damping constant for contact in the normal direction and $v_n$ is the magnitude of the relative velocity between the particle and the surface in the direction perpendicular to the surface. The normal reaction force on the particle acts perpendicular to the flat surface. In our model, we only account for the force on the particle due to the flat substrate and ignore the consequence of the equal and opposite force on the substrate due to the particle. The rationale for ignoring the force on the substrate is that we only consider solid substrates that are much larger and heavier than the particle. Thus, we keep the position of the substrate fixed and do not evolve it when a net force acts on it.

When two spherical particles $i$ and $j$ come into contact, we define the overlap between them as (figure \ref{fig:OverlapDefinition}b)
\begin{equation}
\delta_{ij} = R_i + R_j - |\boldsymbol{X}_i-\boldsymbol{X}_j|,
\label{Overlap}
\end{equation}
where $\boldsymbol{X}_i$ and $\boldsymbol{X}_j$ are the position vector of the centre of mass of particle $i$ and $j$, respectively. When the overlap is positive ($\delta>0$), the magnitude of this force acting each particle is given by
\begin{equation}
F_n=k_n\delta_{ij}+ \gamma_nv_{ij}.
\label{Contact force between two particles}
\end{equation}
Here, $v_{ij}=|(\boldsymbol{v}_i-\boldsymbol{v}_j)\cdot \boldsymbol{n}_{ij}|$ is the component of the relative velocity that points along the direction joining their centre of masses, where $\boldsymbol{n}_{ij}=(\boldsymbol{X}_i-\boldsymbol{X}_j)/|\boldsymbol{X}_i-\boldsymbol{X}_j|$. Both the particles experience an equal normal reaction force, but the force acts in the opposite direction. The direction of the force on the particles points towards their respective centre of mass. For particle $i$, the normal force reaction force vector is $F_n\boldsymbol{n}_{ij}$, whereas for particle $j$, it is $-F_n\boldsymbol{n}_{ij}$, where $F_n$ is given by equation \ref{Contact force between two particles}.

\subsection{Sliding friction}

Sliding friction becomes active when two solid surfaces slide relative to each other. Here, we use the approach proposed by \citet{luding_cohesive_2008} to model sliding friction as arising from a tangential spring that resists relative sliding motion between the two solid surfaces at the contact point. For an initially stationary particle, the spring restores the particle to its initial position when the applied force is below the Coulomb threshold, $F_C=\mu F_n$. When the applied force exceeds this limit, the particle can move continuously while experiencing friction. In the following, we briefly outline the algorithm that we used to implement these principles. A complete description is provided in the paper by \citet{luding_cohesive_2008}.

When a contact is active (\textit{i.e.} positive overlap), we use a linear spring-dashpot model to compute the sliding friction force acting tangential to the contact,
\begin{equation}
    \boldsymbol{F}^0_t(t)=-k_t\boldsymbol{\xi}(t)-\gamma_t\boldsymbol{v}_t(t),
\end{equation}
where $k_t$ is the tangential (or sliding) spring stiffness, $\boldsymbol{\xi}$ is the extension of the tangential spring, $\gamma_t$  is the tangential (or sliding) damping constant, and $\boldsymbol{v}_t$ is the relative velocity between the contacting surfaces at the contact point.

When the restoring force due to the tangential spring is below the Coulomb threshold $(|\boldsymbol{F}_t^0|\leq F_C)$, the particle is in the static regime and the restoring force balances the applied force. In the opposite case when $|\boldsymbol{F}_t^0|> F_C$, the sliding regime becomes active and the magnitude of the friction force is set to $F_C$. By accounting for these two cases, the tangential force applied to the particle can be written as
\begin{equation}
    \boldsymbol{F}_t=\begin{cases}
			\boldsymbol{F}_t^0, & |\boldsymbol{F}_t^0|\leq F_C\\
            F_C\frac{\boldsymbol{F}_t^0}{|\boldsymbol{F}_t^0|}, & |\boldsymbol{F}_t^0|>F_C.
		 \end{cases}
\end{equation}
The extension of the tangential spring is incremented for the next time step, as follows:
\begin{equation}
    \boldsymbol{\xi}(t+\Delta t)=\begin{cases}
        \boldsymbol{\xi}(t)+\boldsymbol{v}_t \Delta t,   &  |\boldsymbol{F}_t^0|\leq F_C\\\\
        -\frac{1}{k_t}\left(F_C\frac{\boldsymbol{F}_t^0}{|\boldsymbol{F}_t^0|}+\gamma_t\boldsymbol{v}_t\right),          & |\boldsymbol{F}_t^0|>F_C.\\
    \end{cases}
\end{equation}
In this expression, when the first condition is met (static regime), the extension of the tangential spring is incremented for the next time step. When the second condition is met (sliding regime), the extension of the spring is defined such that in the next time step, the particle experiences a friction force equal to the Coulomb threshold, $\boldsymbol{F}_t^0(t+\Delta t)\approx F_C\boldsymbol{F}_t^0(t)/|\boldsymbol{F}_t^0(t)|$. 

The torque that arises due to the tangential friction forces is
\begin{equation}
    \boldsymbol{T}_t=(\boldsymbol{X}_C-\boldsymbol{X})\times \boldsymbol{F}_t,
\end{equation}
where $\boldsymbol{X}_C$ is the position vector of the contact point and $\boldsymbol{X}$ is the position vector of the centre of the particle.

\subsection{Rolling Friction}

In addition to sliding friction, objects also experience rolling friction (or rolling resistance) that restrains their rolling motion. In practice, rolling friction can arise due to roughness on the surface of the particle, surface deformation, and asymmetries in the shape of the contact when particles roll.

The model for rolling friction follows a similar line of reasoning as the model for the sliding friction, as explained in detail in \citet{luding_cohesive_2008}. We introduce three new parameters for the rolling friction model: the rolling spring stiffness $k_r$, the rolling damping constant $\gamma_r$, and the coefficient of rolling friction $\mu_r$. To model rolling friction, we follow the same procedure as described above for sliding friction, except for two important differences. First, for the relative velocity at the contact, we use the relative rolling velocity rather than the relative tangential velocity.

Second, the tangential force that is computed is merely a fictive force used to compute the torque. Only the torque is used in Newton's second law for angular motion. The fictive tangential force does not contribute to the equation of motion for the translational motion because rolling resistance only affects the angular motion.

\subsection{Net contact force}

The net contact force acting on a particle when it makes contact with another solid body is obtained by adding the normal reaction, sliding friction, and rolling friction:
\begin{equation}
    \boldsymbol{F}_c=\boldsymbol{F}_n+\boldsymbol{F}_t.
\end{equation}
Rolling friction is not included in the above expression because it only contributes to the torque and not to the net force. The total torque on the particle is given by
\begin{equation}
    \boldsymbol{T}_c=\boldsymbol{T}_t+\boldsymbol{T}_r,
\end{equation}
where $\boldsymbol{T}_r$ is the torque contribution due to rolling friction. Torques due to normal reaction do not enter this expression because normal forces always point towards the centre of the particle for spherical particles and thus do not generate torques.

\subsection {Benchmarking contact forces}

\begin{figure}
  \centerline{\includegraphics[width=1.0\linewidth]{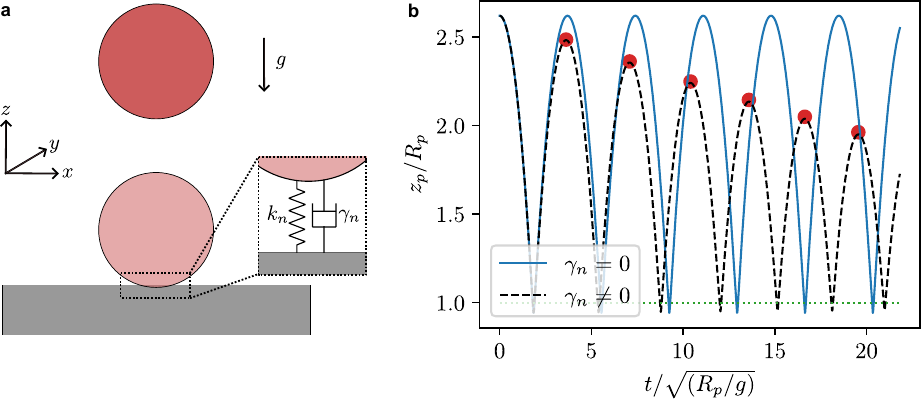}}
  \caption{Particle of radius $R_p$ bouncing off a solid surface. (a) When the spherical solid particle makes contact with the solid surface, the contact force is modelled by a spring-dashpot system. Contact forces are only active when the particle and the surface overlap. (b) Center of mass height of the particle (relative to the top of the solid surface) as a function of time with and without damping. In the absence of damping ($k_n=2.5, \gamma_n=0$), the particle bounces to the same height each time (solid blue line), indicating that energy is conserved. When there is damping ($k_n=2.5, \gamma_n=1.0$), energy is lost during the contact and the maximum height decreases with each contact (dashed black line). The damped rebound height agrees well with the predicted height (Eq.~\ref{Eq. rebound height}, red points). The dotted green horizontal line indicates the solid surface.}
\label{fig:Bouncing}
\end{figure}

We tested the normal reaction force model by analysing the center of mass position of a particle that was released from rest from an initial height $h_0$ (figure \ref{fig:Bouncing}a). In the absence of damping ($\gamma_n=0$), the collision was perfectly elastic, as expected. The sum of the kinetic and gravitational potential energy was conserved. The speed $v_0$ just before impact agrees with the prediction obtained by equating the loss in gravitational potential energy to the gain in kinetic energy, $v_{0}=\sqrt{2gh_0}$, where $g$ is the acceleration due to gravity. Since energy is conserved during the impact, this speed is equal to the speed just after impact. The centre of mass bounced to the same height after every impact (figure \ref{fig:Bouncing}b, solid blue line). Between successive impacts, the centre of mass height varied with time as expected according to the equation of motion, $h(t)=v_0 t-gt^2/2=\sqrt{2gh_0}t-gt^2/2$.

When the collision was damped ($\gamma_n>0$), the centre of mass height decayed with time (figure~\ref{fig:Bouncing}b, dashed line), as expected for situations where energy is lost during a collision. The theoretical rebound height can be predicted from the coefficient of restitution. In the DEM model, the coefficient of restitution is given by~\citep{luding_cohesive_2008}
\begin{equation}
    r=v_{n+1}/v_n=\exp(\eta_n/t_c),
    \label{Eq. Coefficient of restitution}
\end{equation}
where $v_n$ and $v_{n+1}$ are the impact speed after impact number $n$ and $(n+1)$, respectively. The term $\eta_n$ is related to the damping constant according to
$\eta_n = \gamma_n/(2m_p)$ and $t_c=\pi/\omega_c$, where $\omega_c=\sqrt{(k_n/m_p - \eta_n^2)}$. By balancing the kinetic energy just after impact with the potential energy when the particle reaches its maximum height and using Eq.\,\ref{Eq. Coefficient of restitution} to relate the impact speed of the $n^\mathrm{th}$ impact to the impact speed of the previous impact, we obtain a relation for the rebound height reached after the $n^\mathrm{th}$ impact:
\begin{equation}
    h_n = r^{2n} h_0.
    \label{Eq. rebound height}
\end{equation}
The rebound height obtained in our numerical simulations (figure~\ref{fig:Bouncing}b, dashed line) shows excellent agreement with the height predicted by Eq.~\ref{Eq. rebound height} (figure~\ref{fig:Bouncing}b, red points).

\begin{figure}
  \centerline{\includegraphics[width=1.0\linewidth]{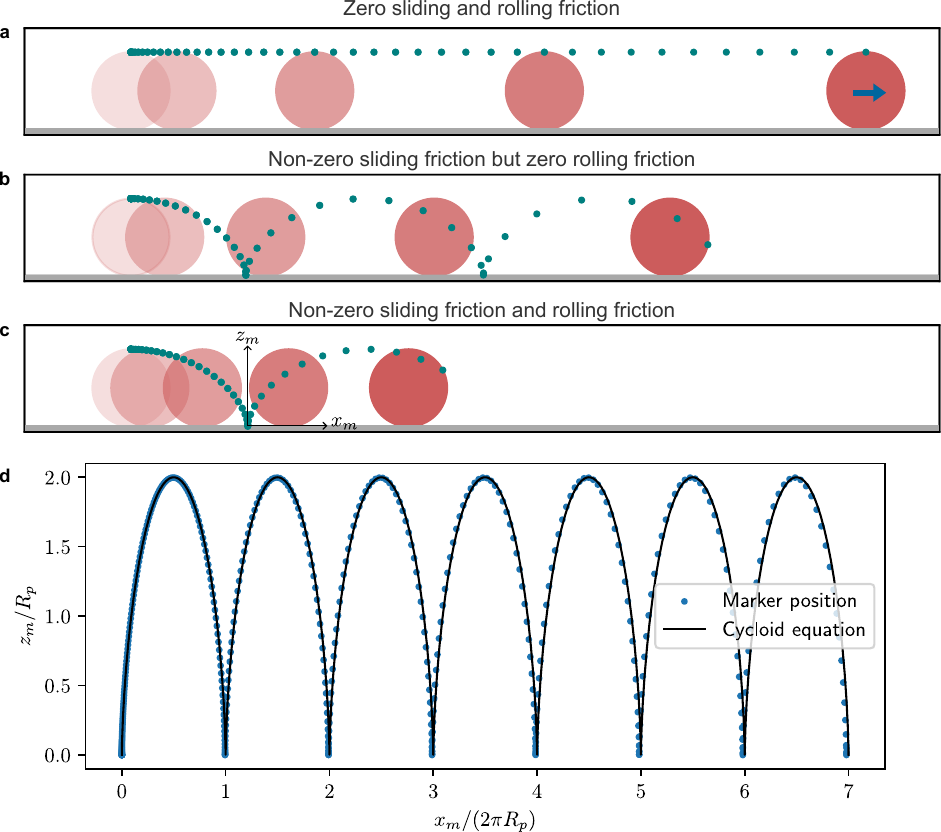}}
  \caption{Motion of a particle down an inclined plane for different friction scenarios: (a) In the absence of sliding friction, the particle slides without any rolling (the green dot on the circumference of the particle is a marker to track the angular motion of the particle); (b) When sliding friction is non-zero, the particle rolls with negligible slip at the contact between the particle and the flat surface; (c) When both sliding and rolling friction are non-zero, the particle rolls slower. (d) Trajectory of the marker point (scatter points) when the particle rolls with negligible slip (\textit{e.g.}, in b and c). The trajectory is described by the cycloid equation (black line), as expected for pure rolling.}
\label{fig:FrictionBenchmarks}
\end{figure}

To benchmark the sliding and rolling friction, we analysed the motion of a sphere down an inclined plane with an inclination angle of $45^\circ$ under gravity (Figure \ref{fig:FrictionBenchmarks}). In the absence of sliding friction ($\mu=0$) and tangential damping ($\gamma_t=0$), the particle slid without any rolling (figure \ref{fig:FrictionBenchmarks}a). The particle moved with a constant acceleration and no energy was lost. When the coefficient of sliding friction was increased to a finite value ($\mu>0$), friction force acted at the contact and opposed sliding. The particle began to roll (figure \ref{fig:FrictionBenchmarks}b). In steady state, the particle displayed an almost pure rolling motion, with a small slip ratio \citep{he_updated_2020}, $s=\omega_p R_p/v_{CM}-1\approx -0.001$, where $\omega_p$ is the angular velocity, $R_p$ is the particle radius, and $v_{CM}$ is the centre of mass velocity of the particle. This expression for the slip ratio quantifies the amount of slip between the particle and the surface at the contact point, with $s\to 0$ corresponding to pure rolling (zero slip) and $s \to-1$ corresponding to pure sliding (\textit{i.e.} when $\omega_p=0)$.

When both rolling friction and sliding friction were non-zero ($\mu>0$, $\mu_r>0$, $\mu_r<\mu$), the particle still rolled, but with a smaller acceleration since energy is now lost due to both sliding and rolling friction (figure \ref{fig:FrictionBenchmarks}c).

When a particle rolls with negligible slip at the contact point, the trajectory traced by a marker point on the circumference of the particle is expected to follow the cycloid equations given by the following parametric equations,
\begin{align}
x_{m}(t) &= R_p[\alpha_p(t)-\sin\alpha(t)], & z_{m}(t) &=R_p[1-\cos\alpha_p(t)]+z_0.
\label{eq.Cycloid}
\end{align}
Here, $x_m$ and $z_m$ are the $x$ and $z$ coordinates (labelled in figure \ref{fig:FrictionBenchmarks}c) traced by the marker point. $z_0$ is the vertical position of the top of the solid surface on which the particle rolls and $\alpha_p$ is the angular position of the marker point with respect to the centre of the particle. The cycloid equation provides an excellent fit to the trajectory traced by the marker point in figure \ref{fig:FrictionBenchmarks}d, as expected for a spherical particle for which the sliding friction exceeds the rolling friction.

\section {Lattice Boltzmann model for fluid dynamics}

In this section, we describe the lattice Boltzmann algorithm used to model the dynamics of the two fluids. Here we focus on an equal-density colour-gradient LBM to model the dynamics of the two liquid components in three-dimensions.

In our colour-gradient model, the two liquids are immiscible and have equal density. The viscosity of the two liquids and the surface tension can be tuned independently. External forces, such as gravity, can be applied to one or both liquids. We outline the method below. For a more detailed description, please refer to previous works by \citet{liu_three-dimensional_2012} and \citet{zhang_rayleighplateau_2022} who have introduced this model in detail.

Each fluid is represented by its respective distribution function, $f^a_i(\boldsymbol{x},t)$ and $f^b_i(\boldsymbol{x},t)$, where $\boldsymbol{x}$ is the position, $t$ is time, the subscript $i$ denotes the discretised velocity, and the superscripts $a$ and $b$ denote fluid $a$ and fluid $b$, respectively. The distribution function gives the number of fluid particles moving at a certain velocity at a given position in space and time. We choose the D3Q19 lattice model to discretise the velocity into 19 possible directions. Details of the lattice velocities $\boldsymbol{c}_i$ and weights $w_i$ associated with each of the discrete velocity directions can be found in \citet{kruger_lattice_2016}.

During each integration time step, the total distribution function, $f_i=f_i^a+f_i^b$, evolves according to
\begin{equation}
    f_i^*(\boldsymbol{x},t)=f_i(\boldsymbol{x},t)+\Omega(\boldsymbol{x},t)+F_i.
    \label{Eq. Collision step}
\end{equation}
Here, $\Omega_i$ is the collision operator that describes how the distribution function relaxes to equilibrium, and $\bar{F}_i$ is a forcing term that incorporates surface tension and external body forces such as gravity. We incorporate these forces using the scheme proposed by \citet{guo_discrete_2002} that accounts for discrete lattice effects and ensures second-order accuracy in space and time. For the collision operator, we use the multiple-relaxation-time (MRT) model to minimise spurious currents and enhance numerical stability. In the MRT model,
\begin{equation}
    \Omega_i(\boldsymbol{x},t)=-\Sigma_j(M^{-1}SM)_{ij}[f_j(\boldsymbol{x},t)-f_j^\mathrm{eq}(\boldsymbol{x},t)].
    \label{collision operator}
\end{equation}
Here, $M$ is the MRT transformation matrix and $S$ is the diagonal relaxation matrix, which encodes information on the local fluid viscosity and is given by \citep{liu_modelling_2020}
\begin{equation}
S=\mathrm{diag}\left[0,\omega,\omega,0,\omega',0,\omega',0,\omega',\omega,\omega,\omega,\omega,\omega,\omega,\omega,\omega',\omega',\omega'\right],
\end{equation}
where $\omega'=8(2-\omega)/(8-\omega)$ and $\omega=\Delta t/\tau$ is the ratio between the integration time step and the relaxation time, $\tau$. 
The relaxation time is related to the dynamic viscosity of the fluid according to $\eta=\rho_f c_s^2(\tau-\Delta t/2)$, where $c_s=\Delta x/(\Delta t\sqrt{3})$ is the speed of sound and $\rho_f$ is the total local fluid density. The total local fluid density is defined as $\rho_f=\rho^a+\rho^b$, where $\rho^a$ and $\rho^b$ are the local density of fluids $a$ and $b$, respectively. Note that the matrix $S$ is a function of position because the dynamic viscosity, and therefore relaxation time, depends on whether a lattice point contains fluid $a$ or $b$ only, or a combination of both (\textit{i.e.} at the interface between the two fluids). The local pressure of the fluid is given by $p=\rho_f c_s^2$.

To distinguish between the two fluids, we define an order parameter as
\begin{equation}
    \phi(\boldsymbol{x},t)=\frac{\rho^a(\boldsymbol{x},t)-\rho^b(\boldsymbol{x},t)}{\rho^a(\boldsymbol{x},t)+\rho^b(\boldsymbol{x},t)}.
\end{equation}
This definition naturally imposes the condition that $-1\leq\phi\leq1$. A value of $\phi=+1$ and $\phi=-1$ corresponds to pure fluid $a$ and $b$, respectively. Values of $-1<\phi<1$ correspond to the interfacial region between the two fluids.

The order parameter is used to obtain the local viscosity and relaxation time in the relaxation matrix. We use the harmonic mean to define the local viscosity as a function of the order parameter as
\begin{equation}
    \frac{1}{\eta(\phi)}=\frac{1+\phi}{2\eta^a}+\frac{1-\phi}{2\eta^b},
\end{equation}
where $\eta^a$ and $\eta^b$ are the dynamic viscosities of fluid $a$ and $b$, respectively. This definition ensures that we recover the viscosity of the pure fluids in regions where there is only a single fluid ($\phi=\pm1$), and assigns a viscosity that lies between that of the two fluids in the interfacial region.

The forcing term $F_i$ in equation \ref{Eq. Collision step} is given by \citep{guo_lattice_2008}
\begin{equation}
F_i = (M^{-1})_{il} \left( \delta_{lk} - \frac{1}{2} S_{lk} \right) M_{kj} \bar{F}_j.
\end{equation}
Here,
\begin{equation}
\bar{F}_i = w_i \left[ \sum\limits_{j}\frac{c_{ij} F^\mathrm{ext}_j}{c_s^2} 
+ \sum\limits_{j,k}\frac{ u_j F^\mathrm{ext}_k (c_{ij} c_{ik} - c_s^2 \delta_{jk})}{c_s^4} \right].
\label{force4}
\end{equation}
In this equation, the indices for the summations run from $j,k=1$ to $3$ to represent the three coordinate dimensions ($x,y,z$). The index $i$ runs from 1 to 19 to denote the discrete velocity directions. The term $F^\mathrm{ext}$ is the applied force, which can be gravity or surface tension between the two fluids. The local macroscopic velocity that enters in the above equation (and is saved to obtain the velocity field) is computed using 
\begin{equation}
\rho_f u_j=\sum_{i=1}^{19} c_{ij}f_i+\frac{\Delta t}{2}F_j,
\end{equation}
where the local fluid density is given by
\begin{equation}
    \rho_f(\boldsymbol{x},t)=\sum_{i=1}^{19} f_i.
\end{equation}
To model surface tension, we use the continuum surface force model to calculate a volume force due to surface tension in the diffuse interface region where the two fluids meet. This force enters as an external force in equation \ref{force4} and is computed using \citep{brackbill_continuum_1992}
\begin{equation}
    F_i^\mathrm{ext}=-\frac{1}{2}\sigma\kappa\partial_i\phi,
\end{equation}
where $\sigma$ is the value of the surface tension, and $\kappa$ is the interface curvature. Note that the expression for the surface tension force gives rise to a Laplace pressure when the interface is curved. When there is no curvature, the surface tension does not contribute to the forcing term.

We apply a recolouring step to promote phase separation and ensure that the two fluids are immiscible using the algorithm proposed by \citet{latva-kokko_diffusion_2005}. In particular, the distribution functions are modified according to
\begin{equation}
    f_i^{a*}(\boldsymbol{x},t)=\frac{\rho^a}{\rho_f}f_i^*(\boldsymbol{x},t)-\beta\frac{\rho^a\rho^b}{\rho_f}w_i(\boldsymbol{c_i}\cdot \boldsymbol{n}),
\end{equation}
\begin{equation}
    f_i^{b*}(\boldsymbol{x},t)=\frac{\rho^b}{\rho_f}f_i^*(\boldsymbol{x},t)+\beta\frac{\rho^a\rho^b}{\rho_f}w_i(\boldsymbol{c_i}\cdot \boldsymbol{n}),
\end{equation}
where $\beta$ is the segregation parameter that controls the interface thickness. We set $\beta=0.7$ to keep the interface narrow and minimise spurious currents \citep{liu_modelling_2020, halliday_lattice_2007}. After the recolouring step, the distribution functions are streamed according to 
\begin{equation}
    f^{a}_i(\boldsymbol{x}+\boldsymbol{c_i}\Delta t, t+\Delta t)=f^{r*}_i(\boldsymbol{x},t).
\end{equation}
\begin{equation}
    f^{b}_i(\boldsymbol{x}+\boldsymbol{c_i}\Delta t, t+\Delta t)=f^{b*}_i(\boldsymbol{x},t).
\end{equation}
The density of each fluid can be calculated as $\rho^{a,b}=\sum_{i}f_i^{a,b}$.

\section{Coupling between fluid and particles}

\begin{figure}
  \centerline{\includegraphics[width=1.0\linewidth]{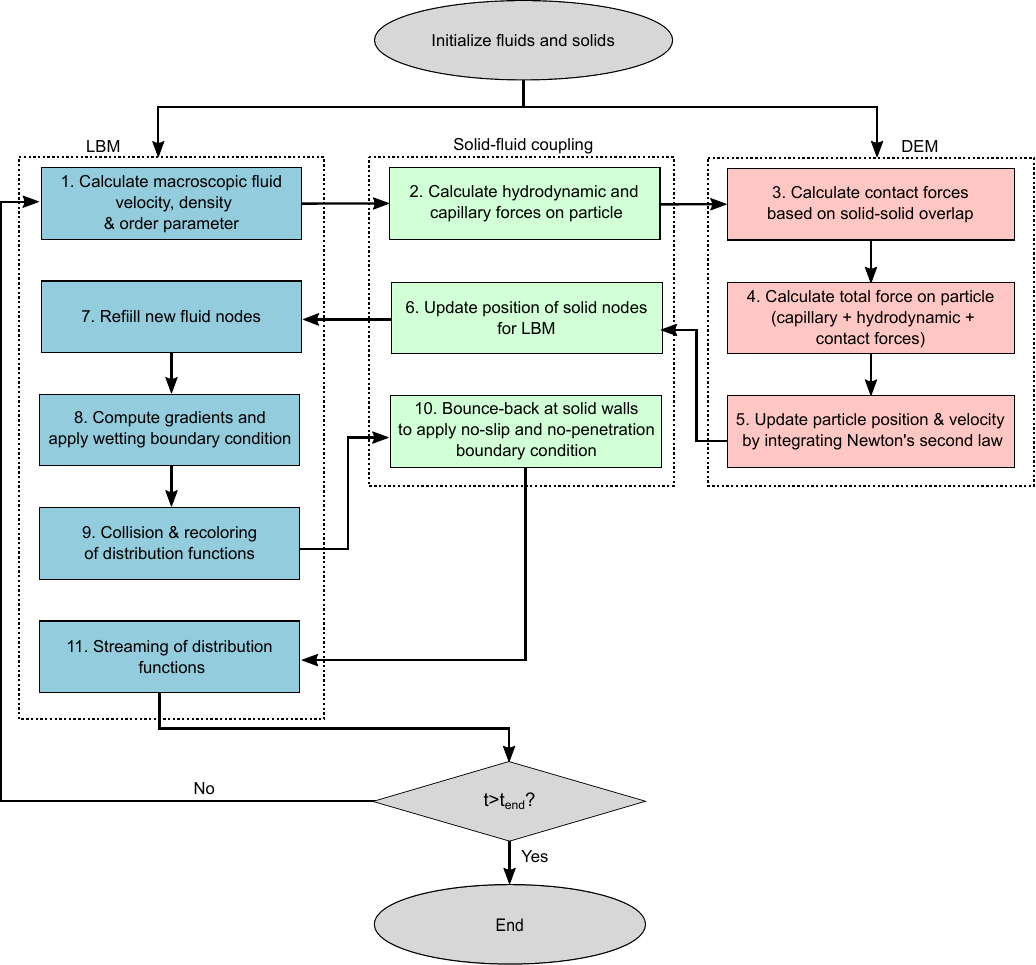}}
  \caption{Flow chart showing the sequence of calculations in the coupled LBM-DEM algorithm.}
\label{fig:Flowchart}
\end{figure}

To couple the two fluids with solid particles, we must account for the no-slip and no-penetration boundary condition at solid walls and consider the transfer of momentum from the fluids to the particle, which gives rise to a hydrodynamic force on the particle. We must also account for the wettability of the solids and compute capillary forces on particles when they are at the interface between the two fluids. In addition to accounting for the above physical interactions, additional technical considerations are also required to reproduce the correct physical behaviour. First, fluid nodes that emerge at the receding side of a solid particle must be refilled with fluid. Second, we have to ensure that there is always at least one fluid node separating two solid particles for the lattice Boltzmann algorithm to be stable. In this section, we describe how we account for all these forces and technical considerations. The flow chart in Figure \ref{fig:Flowchart} summarises the sequence of calculations performed by the LBM and DEM algorithms, and the coupling between them.

\subsection{No-slip and no-penetration at solid boundaries}

To achieve the no-slip and no-penetration boundary conditions at the solid wall, we modify the distribution functions of fluids $a$ and $b$ using the half-way bounce-back scheme as follows:
\begin{equation}
f^a_i(\boldsymbol{x}_f,t)=f^a_{\bar{i}}(\boldsymbol{x}_b,t)+6w_i\rho^a(\boldsymbol{x}_f,t)\frac{\boldsymbol{u}_w(\boldsymbol{x}_w,t)\cdot\boldsymbol{c}_i}{c_s^2}, 
\end{equation}
\begin{equation}
f^b_i(\boldsymbol{x}_f,t)=f^b_{\bar{i}}(\boldsymbol{x}_b,t)+6w_i\rho^b(\boldsymbol{x}_f,t)\frac{\boldsymbol{u}_w(\boldsymbol{x}_w,t)\cdot\boldsymbol{c}_i}{c_s^2}, 
\end{equation}
where $\boldsymbol{u}_w$ is the velocity of the particle wall and the subscript $\bar{i}$ denotes the discrete velocity direction opposite to $i$. This update is performed after the streaming step. After streaming, the distribution function will have propagated into the solid. $\boldsymbol{x}_b$ are locations of solid nodes that neighbour at least one fluid node. When computing the wall velocity, both the translational and rotational motion are considered as follows:
\begin{equation}
\boldsymbol{u}_w=\boldsymbol{u}_p+\boldsymbol{\omega}_p\times (\boldsymbol{x}_w-\boldsymbol{x}_p),
\end{equation}
where $\boldsymbol{x_p}$ and $\boldsymbol{u_p}$ is the position and velocity of the centre of mass of the particle, respectively. On fixed solid substrate, $\boldsymbol{u}_w=0$ since the solid is stationary.

\subsection{Transferring momentum from fluid to particle}

When fictive fluid particles bounce back on the surface of solid particles, they transfer momentum to the particle. We use the momentum exchange method to obtain the momentum transferred to the solid particle \citep{wen_galilean_2014}. Other methods that can used for the force coupling between the fluid and the particle include the immersed boundary method \citep{peskin_numerical_1977} or the stress integration method. Here, we choose the momentum exchange method because of its simplicity and its intuitive physical interpretation. With this method, momentum is transferred along each of the 19 discrete velocity directions. The momentum transferred to the particle along the $i$ direction per time step can be obtained from the distribution functions according to
\begin{equation}
    \Delta \boldsymbol{p}(\boldsymbol{x}_w)= \sum_{i=1}^{19}\left[ (\boldsymbol{c}_{\bar{i}}-\boldsymbol{u}_w)f_{\bar{i}}^{*}-(\boldsymbol{c}_i-\boldsymbol{u}_w)f_i \right],
\end{equation}
where $f^*$ denotes the distribution function after bounce-back and $f$ denotes the distribution function before bounce-back. The summation runs over all discrete velocity directions. The change of momentum per time step gives the force on the particle. The total hydrodynamic force and torque on the particle is obtained by summing the local force at all solid wall nodes,
\begin{equation}
    \boldsymbol{F}_\eta=\sum_\mathrm{Wall} \Delta \boldsymbol{p}(\boldsymbol{x}_w),
    \label{hydro1}
\end{equation}
\begin{equation}
    \boldsymbol{T}_\eta=\sum_\mathrm{Wall} (\boldsymbol{x}_w-\boldsymbol{x}_p)\times \Delta \boldsymbol{p}(\boldsymbol{x}_w).
    \label{hydro2}
\end{equation}
Here, the summation runs over all solid wall nodes.

\subsection{Refilling of new fluid nodes that emerge on the receding side of a particle}

When particles move, nodes that were previously in the solid region may be converted to fluid nodes. The distribution functions at these fresh fluid nodes must be assigned since we do not evolve the distribution function inside the solid. To assign distribution functions at these fresh fluid nodes, we calculate the corresponding equilibrium distribution function from density of the neighbouring fluid nodes and the macroscopic velocity of the solid wall as follows \citep{caiazzo_analysis_2008}:
\begin{equation}   f^{a,eq}_i(\boldsymbol{x}_n)=\rho^a(\boldsymbol{x}_n)w_i\left(1+\frac{\boldsymbol{u}(\boldsymbol{x}_n)\cdot\boldsymbol{c}_i}{c_s^2}+\frac{(\boldsymbol{u}(\boldsymbol{x}_n)\cdot\boldsymbol{c}_i)^2}{2c_s^4}-\frac{\boldsymbol{u}(\boldsymbol{x}_n)\cdot\boldsymbol{u}(\boldsymbol{x}_n)}{2c_s^2}\right),
\end{equation}
\begin{equation} f^{b,eq}_i(\boldsymbol{x}_n)=\rho^b(\boldsymbol{x}_n)w_i\left(1+\frac{\boldsymbol{u}(\boldsymbol{x}_n)\cdot\boldsymbol{c}_i}{c_s^2}+\frac{(\boldsymbol{u}(\boldsymbol{x}_n)\cdot\boldsymbol{c}_i)^2}{2c_s^4}-\frac{\boldsymbol{u}(\boldsymbol{x}_n)\cdot\boldsymbol{u}(\boldsymbol{x}_n)}{2c_s^2}\right),
\end{equation}
where $\boldsymbol{x}_n$ denotes the position of a fluid node that neighbours a solid particle node. The local density at the fresh fluid node is computed by taking an average of the local density at all neighbouring fluid nodes that lie in any of the 19 directions of the D3Q19 velocity set,
\begin{equation}
    \rho^{a,b}(\boldsymbol{x}_n)=\frac{1}{N_f}\sum^{N_f}_{i \ \in \ \mathrm{ Fluid}} \rho^{a,b}(\boldsymbol{x}_n+\boldsymbol{c}_i\Delta t,t),
\end{equation}
where $N_f$ is the number of fluid nodes surrounding the fresh fluid node that is being considered. The summation excludes neighbouring solid nodes and freshly created fluid nodes. This averaging procedure does not lead to local instability in density because the two fluids have equal density in our model. The macroscopic velocity is obtained from the particle velocity as
\begin{equation}
    \boldsymbol{u}(\boldsymbol{x}_n)=\boldsymbol{u}_p+\boldsymbol{\omega}_p\times(\boldsymbol{x}_n-\boldsymbol{x}_p).
\end{equation}
Here $\boldsymbol{v}_p$ is the particle centre of mass velocity. Note that while the fluid is constrained to a lattice grid, the position of the particle is continuous and is not set by the fluid grid.

\subsection{Benchmarking hydrodynamic force}

\begin{figure}
  \centerline{\includegraphics[width=1.0\linewidth]{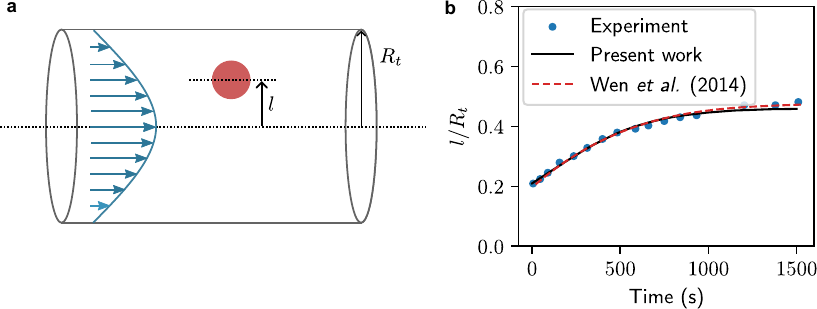}}
  \caption{Benchmarking the hydrodynamic force by investigating the migration of a spherical particle in a Poiseuille flow in a cylindrical tube (Segr\`e-Silberberg effect). (a) Schematic of the simulation setup. The Poiseuille flow was generated by applying a pressure inlet/outlet boundary condition. (b) Motion of the particle perpendicular to the cylindrical axis as a function of time. Our numerical results (black line) agree well with previous experimental results by \citet{karnis_flow_1966} and previous numerical results by \citet{wen_galilean_2014}.}
\label{fig:Hydrodynamic benchmark}
\end{figure}

To validate the accuracy of the hydrodynamic force calculation, we simulated the migration of a neutrally buoyant rigid sphere in a cylindrical Poiseuille flow, as schematically illustrated in figure~\ref{fig:Hydrodynamic benchmark}a. In this simulation, the sphere is expected to eventually reach an equilibrium position within the cylindrical tube, a phenomenon known as the Segrè–Silberberg effect \citep{karnis_flow_1966}.

For this benchmark, we used the following parameter values in simulation units: tube radius $R_t=40$, particle radius $R_p=12.2$, fluid density $\rho_f=1.05$, fluid kinematic viscosity $\eta/\rho_f=0.457$, and average velocity at inlet $v_\mathrm{in}=0.0113$. Initially, the particle was positioned at $l/R_t = 0.21$, where $l$ is the distance from the central axis of the tube as shown in figure~\ref{fig:Hydrodynamic benchmark}a). These values were chosen to match the experimental parameters used by \citet{karnis_flow_1966} to allow a direct comparison with their experiments. The corresponding experimental parameters are: $R_t = 2\,\mathrm{mm}$, $R_p = 0.61\,\mathrm{mm}$, $\rho_f=1.05 \times 10^3\,\mathrm{kg/m^3}$, $\eta/\rho_f=1.14 \times 10^{-5}\,\mathrm{m^2/s}$, and $v_\mathrm{in}=5.6\,\mathrm{mm/s}$. 

At both the inlet and outlet of the tube, we applied pressure boundary conditions applied using the anti-bounce-back scheme to modify the distribution functions as follows \citep{kruger_lattice_2016}:
\begin{equation}
f_{\bar{i}}(\boldsymbol{x}_w, t + \Delta t) = -f_i^*(\boldsymbol{x}_w, t) + \frac{2w_i}{c_s^2} 
\left[ p_w + {\rho_w}{c_s^2} \left(\frac{(\boldsymbol{c}_i \cdot \boldsymbol{u}_w)^2}{2c_s^4} - \frac{u_w^2}{2c_s^2} \right)\right].
\end{equation}
Here, $p_w$, $\rho_w$ and $\boldsymbol{u}_w$ correspond to the fluid pressure, density, and velocity at the inlet/outlet boundaries, respectively. As shown in figure \ref{fig:Hydrodynamic benchmark}b, our numerical results show good agreement with previous experimental~\citep{karnis_flow_1966} and numerical~\citep{wen_galilean_2014} results. Therefore, errors that propagate into the hydrodynamic force calculation due to the refilling procedure and the staircase approximation of the particle shape in the bounce-back procedure do not affect the particle dynamics significantly.

\subsection{Wetting boundary condition}

When the interface between the two fluids contacts a solid surface (either the particle or the flat solid substrate), it forms a contact angle that depends on the surface energy of the solid. We use a geometric wetting boundary condition that allows us to independently tune the contact angle. The order parameter gradient at the fluid nodes next to a solid boundary within the diffuse interface region is modified such that it results in the prescribed contact angle \citep{xu_lattice_2017, akai_wetting_2018}. Two possible vectors, $\boldsymbol{n}_+$ and $\boldsymbol{n}_-$ are computed for the modified gradient direction,
\begin{equation}
    \boldsymbol{n}_{\pm}=\left(\cos\pm \theta_p-\frac{\sin \pm \theta_p \cos \alpha^*}{\sin\alpha^*}\right)\boldsymbol{n}_s+\frac{\sin \pm \theta_p}{\sin\alpha^*}\boldsymbol{n}^*.
\end{equation}
Here, $\boldsymbol{n}_s$ is the unit normal vector that points outwards from the surface of the solid, $\alpha^*=\cos^{-1}(\boldsymbol{n}_s\cdot\boldsymbol{n}^*)$, and $\boldsymbol{n}^*$ is the initial direction of the gradient of the order parameter calculated before applying the correction. Here $\theta_p$ is the prescribed contact angle. The vector that is closest to $\boldsymbol{n}^*$ is chosen as the gradient direction. The direction of the gradient is then updated (while keeping the magnitude constant) to obtain the prescribed contact angle.

\begin{figure}
  \centerline{\includegraphics[width=1.0\linewidth]{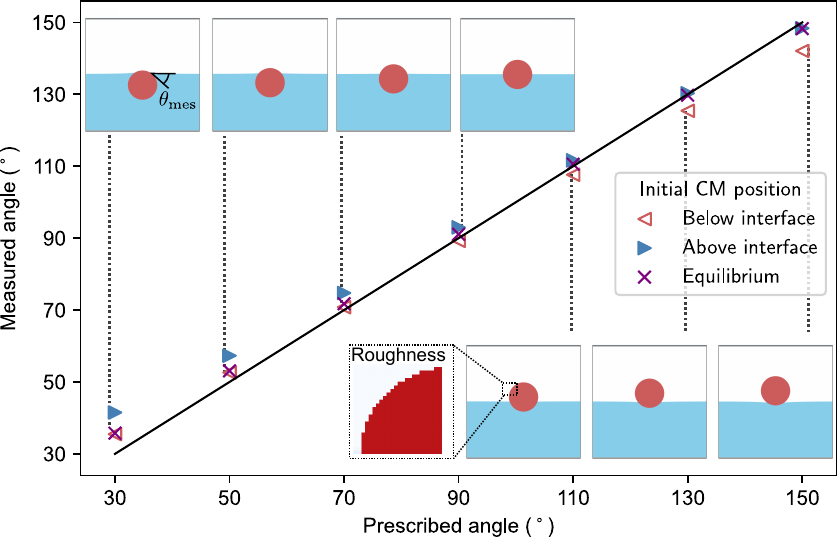}}
  \caption{Benchmark of the wetting boundary condition between the particle and the fluids by varying the prescribed contact angle and measuring the corresponding contact angle obtained in the simulations. The snapshots show the final positions of the particle relative to the interface between the two fluids for each contact angle. The line denotes the trend that should be obtained if there is perfect agreement between the prescribed and measured angles. The data points lie close to the line, showing that the wetting boundary condition accurately models contact angles between $30^\circ$ and $150^\circ$ (inclusive). For these simulations, the domain size is $200\times200\times200$ and $R_p=25$. The viscosities of the two liquids are equal (relaxation times, $\tau_a=\tau_b=1.0$) and density the fluid density is $\rho_f=0.05$. The interfacial tension between the two fluids is $\sigma=0.02$. The filled triangles, crosses, and empty triangles correspond to when the particle was initialised above, at, and below its predicted equilibrium position relative to the interface to demonstrate contact angle hysteresis. Contact angle hysteresis arises because the fluid sees the particle as being rough due to the lattice Boltzmann grid. The image processing procedure to obtain the snapshots shown here is described in Supplementary Figure 1.}
\label{fig:WettingBenchmark}
\end{figure}

We benchmarked the wetting boundary condition by investigating the equilibrium position of a particle at a liquid-fluid interface for a range of prescribed contact angles, as shown in figure \ref{fig:WettingBenchmark}. For this benchmark, the viscosity of the two fluids was set to be equal. The fluid parameters were chosen such that the Ohnesorge number, which characterises the ratio between the interfacial forces and the viscous forces, was $\mathrm{Oh}=\eta/\sqrt{\rho_f R_p\sigma}=0.075$. Initially, the particle was positioned exactly halfway across the liquid-fluid interface. Gravity was neglected. As the simulation evolved, the capillary force shifted the particle upward (when $\theta_p>90^\circ$) or downward (when $\theta_p<90^\circ$) until the interface relaxed back to its equilibrium shape and there was no net capillary force on the particle. By measuring the final position reached by the centre of the particle, we can obtain the actual measured contact angle between the particle and the lower fluid and compare it with the prescribed contact angle to test the implementation of the wetting boundary condition. Based on circular geometry, the measured angle (in radians) is related to the centre of the particle according to
\begin{equation}
\theta_\mathrm{mes}=\frac{\pi}{2}-\arcsin\left(\frac{z_p-z_\mathrm{int}}{R_p}\right),
\end{equation}
where $z_p$ is the final vertical position of the particle's centre of mass and $z_\mathrm{int}$ is the position of the liquid-fluid interface.

We obtain a good agreement between the prescribed contact angle and the measured contact angle (Figure \ref{fig:WettingBenchmark}), especially for contact angles $>30^\circ$ when the particle is initialised close to its predicted equilibrium position (in which case the difference between the predicted and measured angles $\approx 6\%$). These results do not depend on the viscosity ratio between the two fluids as long as the particle is allowed to reach its final steady-state position.

We also observed contact angle hysteresis \citep{butt_contact_2022} when the particle was initialised away from its equilibrium position. When the particle was initialised above the interface such that it had to move downward to reach its final position, the measured angle corresponds to the advancing contact angle, as shown in figure \ref{fig:WettingBenchmark} (blue filled triangles). In contrast, when the particle was initialised below the interface such that it had to move upward to reach its final position, the measured angle was lower and corresponds to the receding contact angle, as shown in figure \ref{fig:WettingBenchmark} (empty red triangles). Contact angle hysteresis is expected because the particle does not appear perfectly smooth from the LBM perspective. Instead, it appears rough because the solid nodes are confined to a lattice grid. Roughness leads to contact angle hysteresis and prevents the particle from reaching the equilibrium position that it would have reached if it had been perfectly smooth.

\subsection{Capillary force}

When particles are at the liquid-fluid interface, they experience capillary force due to the action of surface tension at the three-phase contact line. The capillary force depends on several factors, including the particle shape and size, the interfacial tension between the two fluids, and the contact angle between the particle and the two fluids.

In principle, the capillary force can be obtained by integrating the surface tension vector along the three-phase contact line,
\begin{equation}
\boldsymbol{F}_\sigma=\oint_\mathrm{CL}\boldsymbol{\sigma}\cdot\mathrm{d}\boldsymbol{l},
\end{equation}
where $\boldsymbol{l}$ is the line element and the integral is performed around the contact line. However, performing this integral is not straightforward in LBM. Since we use a diffuse interface model, this integral has to be modified to account for the fact that the contact line is not a sharp line but a ring with a finite thickness. Furthermore, the integral must be discretised and converted into a summation over a set of Lagrangian points spanning the surface of the particle.  In this work, we use the capillary force model by \citet{zhang_rayleighplateau_2022}. By taking into account the above considerations, \citet{zhang_rayleighplateau_2022} proposed that the capillary force on a particle can be computed as
\begin{equation}
    \boldsymbol{F}_\sigma=\sum_i\frac{9}{2}k \beta[1-\phi(\boldsymbol{x_i})^2]\Delta A_i \sin\theta \ \boldsymbol{m}(\boldsymbol{x}_i).
\end{equation}
Here, the summation runs across all Lagrangian points on the surface of the particle that lie within the diffuse interface region where $-1<\phi<1$, $k$ is a lattice-dependent geometric constant that links the segregation parameter $\beta$ to the equilibrium interface thickness. For the D3Q19 discrete velocity set used in our method, $k\approx0.134$ \citep{zhang_rayleighplateau_2022, riaud_lattice-boltzmann_2014}. The term $\Delta A_i=R_p^2\sin\alpha_i\Delta\alpha\Delta\varphi$ is the area element on the spherical particle surface at the Lagrangian point $i$, where $\alpha_i$ is the polar angle of the Lagrangian point and $\varphi$ is the azimuthal angle. Throughout this paper, we set $\Delta\alpha=\Delta \varphi=2^\circ$. The vector $\boldsymbol{m}(\boldsymbol{x}_i)=(\boldsymbol{n}(\boldsymbol{x}_i)\times\boldsymbol{n}_s(\boldsymbol{x}_i))\times \boldsymbol{n}(\boldsymbol{x}_i)$  is the unit vector tangential to the liquid--fluid interface at the Lagrange point $\boldsymbol{x}_i$.

When the capillary force is distributed unequally around the particle, for example, due to a difference in contact angle on different sides of the particle, this leads to a capillary torque \citep{naga_capillary_2021}. This consideration is implicitly embedded in the model.

\subsection{Benchmarking capillary force}

Next, we focus on benchmarking the capillary force when a particle is at the liquid interface. The capillary force plays a key role in understanding how drops interact with particles on surfaces, which we will discuss in the next section.

\begin{figure}
  \centerline{\includegraphics[width=1.0\linewidth]{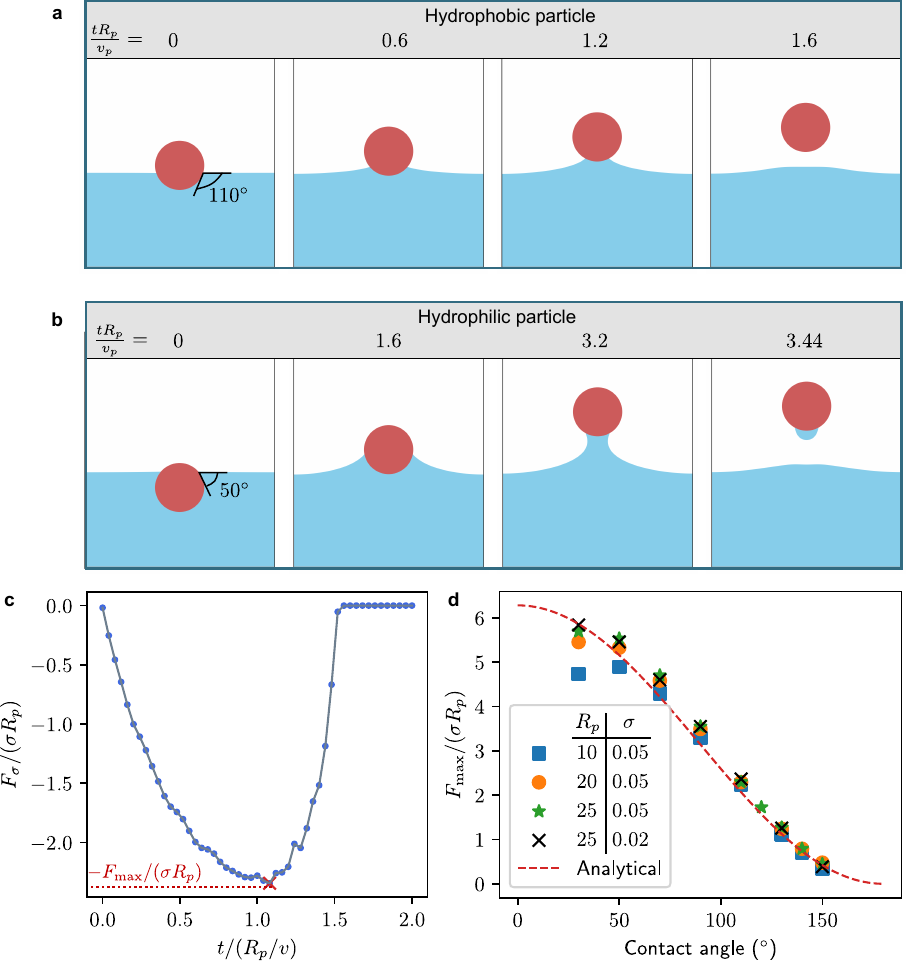}}
  \caption{Benchmarking the capillary force by detaching a particle from a liquid--fluid interface. (a) Simulation snapshots show the detachment process of a hydrophobic particle ($R_p=25$ lattice units, $\theta_p=110^\circ$) as the particle moves upward at a constant velocity ($v_p=0.01$). (b) Detachment of a hydrophilic particle ($R_p=25$, $\theta_p=50^\circ$) from an interface. A small amount of liquid residue remains on the particle after the detachment. (c) Vertical component of the normalised capillary force on the particle against normalised time during the detachment. This force curve corresponds to the snapshots shown in (a) for $\theta_p=110^\circ$. On the $x-$axis, time is normalised by the time taken for the particle to move a distance of 1 particle radius, $R_p/v_p$. The minimum of the force curve is given by $F_\mathrm{max}$. (d) $F_\mathrm{max}$ against $\theta_p$ for contact angles between $30^\circ$ and $150^\circ$. The results are in good agreement with the analytical prediction (dashed red line), in particular when using large particles ($R_p=25$).}
\label{fig:DetachmentBenchmark}
\end{figure}

We benchmarked the capillary force by measuring the force required to detach a particle from a liquid--fluid interface, as shown in figure \ref{fig:DetachmentBenchmark}. For this benchmark, we initialised the particle halfway across the interface and allowed it to reach its equilibrium position, which is a function of the contact angle. We then moved the particle upward at constant velocity and measured the capillary force due to the action of surface tension at the three-phase contact line (figure \ref{fig:DetachmentBenchmark}c). When the particle was hydrophobic ($\theta_p>90^\circ$), it detached cleanly from the interface (figure \ref{fig:DetachmentBenchmark}a), and no liquid residue was left on the particle. In contrast, with hydrophilic particles ($\theta_p<90^\circ$), the interface became unstable and ruptured, leaving a liquid residue on the particle (figure \ref{fig:DetachmentBenchmark}b).

During the detachment, the magnitude of the capillary force increased (figure \ref{fig:DetachmentBenchmark}c) and reached a maximum. The measured maximum force can be compared with the following analytical prediction \citep{scheludko_measurement_1975}:
\begin{equation}
    F_\mathrm{max}=2\pi \sigma R_p\cos^2\left(\frac{\theta_p}{2}\right).
    \label{equation:detachment force}
\end{equation}
Here, this expression assumes that the contact angle maintains a constant value $\theta_p$ as the three-phase contact line slides on the particle. It also neglects the effects arising from the rupture of the capillary bridge. To test our capillary force model, we compared the maximum measured forces with the predictions given by equation \ref{equation:detachment force} for a range of contact angles. Our simulation results agree well with the analytical prediction for $30^\circ \leq \theta_p \leq 150^\circ$, as shown in figure \ref{fig:DetachmentBenchmark}d. For small contact angles ($\theta_p=30^\circ$), the error in the capillary force can be as large as 30\% for small particles with radius $10$ lattice units. The error can be reduced to within 5\% by increasing the particle radius to $25$ lattice units. Two factors contribute to this error. First, the prescribed contact angle that we use in the analytical prediction (equation \ref{equation:detachment force}) is not exactly equal to the actual contact angle on the particle, as shown in figure \ref{fig:WettingBenchmark} where discrepancies can be as large as $\approx 16\%$ for the worst cases. Second, our capillary force model computes the capillary force assuming that the particle is a perfect sphere, but from the LBM perspective, the particle is jagged because it must conform to the discrete lattice grid. This explains why the agreement is worse for smaller particles in figure \ref{fig:DetachmentBenchmark}d. For small particles the assumption of a perfect sphere becomes less realistic due to the jagged edges becoming more significant. Therefore, when choosing the particle size, we must make a compromise between the computational cost required and the accuracy of the capillary force model.

\subsection {Other technical considerations}

In LBM, there must always be at least one fluid node separating two particles or a particle and a solid substrate for the hydrodynamic force calculated by the momentum exchange method to be accurate~\citep{zhang_rayleighplateau_2022,jiang_coupled_2022}. However, DEM requires a finite overlap between solids for contact forces to become active. To harmonise the differing requirements of LBM and DEM, we use a DEM particle radius (which we call the mechanical contact radius, $R_c$) that is larger than the LBM radius (which we call the effective hydrodynamic radius, $R_p$), following the approach proposed by \citet{fei_coupled_2023, jiang_coupled_2022} and \citet{boutt_coupled_2011}. With this approach, there can still be contact from the perspective of DEM while ensuring that there is always at least one fluid node separating the solid nodes between two particles or between a particle and the solid substrate. The difference between the mechanical and hydrodynamic radii, $\delta R=R_c-R_p$, is set to one lattice unit in our simulations. As suggested by \citet{jiang_coupled_2022}, this approach qualitatively mimics a similar effect to lubrication forces, since the contact forces start to become active when there is still a thin liquid film separating the hydrodynamic radii of the two solid surfaces. Importantly, for the particle removal scenarios that we will focus on in Section\,\ref{section: Particle removal}, a quantitative treatment of lubrication forces is not necessary because there is direct contact between the particle and the surface, as shown in previous experiments with hydrophobic particles on hydrophobic surfaces~\citep{naga_thesis_2021}.

To resolve the contact force dynamics, the integration time step $\Delta t$ must be smaller than the duration $t_c$ of a collision between a particle and a solid surface or between two particles. The duration of a collision is given by
\begin{equation}
  t_c=\frac{\pi}{\omega_c},
  \label{collision duration}
\end{equation}
where $\omega_c=\sqrt{\frac{k_n}{m_p}-\left(\frac{\gamma_n}{2m_p}\right)^2}$ is the angular frequency of the damped oscillations that arise when a particle collides with a solid substrate. For the collision between two particles, $m_p$ must be replaced by the reduced mass of the particles, $m_{ij}=m_im_j/(m_i+m_j)$. The term within the square root of the denominator must be positive for $t_c$ to be a real number. This imposes a constraint on the maximum damping constant that can be used for a given spring constant and particle mass, $\gamma_n<2\sqrt{k_nm_p}$.  In LBM, the integration time step is typically set to unity. The integration time step for the DEM model can also be set to unity as long as $\Delta t\ll t_c$. Throughout this paper, we use $\Delta t \approx t_c/45$. 

\section{Investigating particle removal by drops}\label{section: Particle removal}

\begin{table}
  \begin{center}
\def~{\hphantom{0}}
  \begin{tabular}{lcccc}
      Property  & Symbol   &   Value in simulation units & Value in SI units\\[3pt]
      \hline
       Time unit   & $\Delta t$ & 1 & $2.53\times10^{-7}$\,s\\
       Length unit   & $\Delta x$ & 1 & $1.00\times10^{-5}$\,m\\
       Fluid density & $\rho_f$ & 0.05 & $4.63$\,$\mathrm{kg\,m^{-3}}$\\
       Surface tension between two fluids & $\sigma$ & 0.01--0.05& $14.5\times10^{-3}$--$72.3\times10^{-3}$\,$\mathrm{kg\,m\,s^{-2}}$\\
       Dynamic viscosity of fluid $a$ (air) & $\eta_a$ & 0.00273 & $10.0\times10^{-4}$\,$\mathrm{kg\,m^{-1}\,s^{-1}}$  \\
       Dynamic viscosity of fluid $b$ (drop) & $\eta_b$ & 0.0273 & $10.0\times10^{-3}$\,$\mathrm{kg\,m^{-1}\,s^{-1}}$\\
       Fluid--particle contact angle & $\theta_p$ & $90^\circ$--$95^\circ$ & $90^\circ$--$95^\circ$\\
       Fluid--substrate contact angle & $\theta_s$ & $90^\circ$--$110^\circ$ & $90^\circ$--$110^\circ$\\
       Particle radius   & $R_p$ & 10& $100\times10^{-6}$\,$\mathrm{m}$ \\
       Gravitational force on particle (in air) & $g_p$ & 0.52--5.23& $7.57\times10^{-6}$--$7.57\times10^{-5}$\,$\mathrm{kg\,m\,s^{-2}}$ \\
       Total horizontal force on drop & $g_d$ & 0.26--1.57 & $3.78\times10^{-6}$--$2.27\times10^{-5}$\,$\mathrm{kg\,m\,s^{-2}}$\\
       Particle density & $\rho_p$ & 0.125 & $11.6$\,$\mathrm{kg\,m^{-3}}$\\
       Normal spring stiffness & $k_n$ & 2.5 & 3.61\,$\mathrm{kg\,s^{-2}}$\\
       Sliding spring stiffness & $k_t$ & 1.0& 1.45\,$\mathrm{kg\,s^{-2}}$ \\
       Rolling spring stiffness & $k_r$ & 0.5& 0.72\,$\mathrm{kg\,s^{-2}}$ \\
       Sliding friction coefficient & $\mu$ & 0--1.0&0--1.0\\
       Rolling friction coefficient & $\mu_r$ & 0--1.0&0--1.0\\
       Normal damping constant & $\gamma_n$ & 0--0.2& 0--$7.31\times10^{-8}$\,$\mathrm{kg\,s^{-1}}$\\
       Sliding damping constant & $\gamma_t$ & 0--1.0& 0--$3.66\times10^{-7}$\,$\mathrm{kg\,s^{-1}}$ \\
       Rolling damping constant & $\gamma_r$ & 0--1.0& 0--$3.66\times10^{-7}$\,$\mathrm{kg\,s^{-1}}$ \\
\hline
  \end{tabular}
              \caption{Description and values of the parameters used when simulating the collision between a drop and a particle on a flat solid surface. The conversion factor for the length scale is chosen such that the particle radius is 100\,$\mathrm{\mu}$m, and the conversion factors for the mass and time are chosen such that the surface tension matches the surface tension between air and water (72\,mN/m) and the dynamic viscosity of the drop matches that of water (1\,mPa\,s) for some of the simulations. All other quantities can be converted from the conversion factors for length, mass, and time because the SI units of these quantities are combinations of these three quantities. The viscosity ratio between the two fluids is 10. The contact angles are chosen to match the experimental parameters in \citep{naga_how_2021}. When comparing with experiments, the gravitational force on the particle was chosen such that the friction force on the particle has the same order of magnitude as the capillary force between the drop and the particle, as is the case in \citep{naga_how_2021}. Varying the damping constants $\gamma_n$, $\gamma_t$, and $\gamma_r$ in the range indicated above did not noticeably affect the results. Therefore, when conducting the systematic investigations in figures\,\ref{fig:varymu}, \ref{fig:vary_drop_vol}, \ref{fig:vary_drop_force}, they were set to zero.} 
  \label{tab:parameters}
  \end{center}
\end{table}

Self-cleaning surfaces are solid surfaces that can be easily cleaned by the passage of liquid drops. These surfaces were originally inspired by the lotus leaf. Despite being exposed to dust and dirt, the lotus leaf remains clean because raindrops and fog can easily capture dust/dirt particles while rolling off the leaf~\citep{barthlott_purity_1997,geyer_when_2020}. When a drop slides across a surface and collides with a particle, several forces are involved, including sliding and rolling friction between the particle and the solid surface, and capillary and hydrodynamic forces between the particle and the drop. In this section, we demonstrate how all the different ingredients of our coupled LBM--DEM method can be combined to study the mechanism of particle removal by drops on a flat solid surface in three dimensions.

\subsection{Simulation setup and parameters}

To investigate how a drop removes a particle from a surface, we initialised the drop as a hemisphere on a horizontal solid surface and positioned the particle on the surface in front of the drop. For all the simulations presented in this section, the domain size was 4\,mm$\times$1.81\,mm$\times$ 7.1\,mm ($400\times181\times71$ lattice units) and the particle radius was $R_p=100$\,$\mathrm{\mu}$m (10 lattice units). The drop radius was $R_d=5R_p$, except in section\,\ref{vary size and speed} where we varied the radius from $2R_p$ to $6R_p$.

To move the drop along the horizontal surface, we applied a horizontal force $f_x$ per unit volume to the drop along $x$. On the particle, we applied a downward force $g_p$ (henceforth called the gravitational force on the particle) to account for the combined effects of gravity and intermolecular adhesion forces when the particle is in contact with the surface. Our model also accounts for the reduction in the effective gravitational force on the particle due to buoyancy when the particle enters the liquid drop. Periodic boundary conditions were applied along all three Cartesian directions. 

In this paper, we systematically investigate the effects of varying the coefficients of friction between the particle and the surface, the radius of the drop, and the force applied to move the drop, expressed in terms of the dimensionless Bond number, $\mathrm{Bo}=f_xR_d/\sigma$. We focus on hydrophobic particles ($90^\circ \leq \theta_p \leq 95^\circ$) on hydrophobic surfaces ($90^\circ \leq \theta_s \leq 110^\circ$). The full list of parameters is provided in Table\,\ref{tab:parameters}.

Our simulations correspond to Bond numbers in the range of $0.16\leq\mathrm{Bo}\leq1.44$. Within this range, the capillary number ($\mathrm{Ca}=\eta_bU/\sigma$, where $\eta_b$ is the dynamic drop viscosity, $\sigma$ is the surface tension, and $U$ is the drop centre of mass velocity) is of the order of $10^{-2}$ or lower, indicating that the capillary force on the particle dominates the viscous force. Furthermore, the particle Reynolds number ($\mathrm{Re}=\rho_f R_p U/\eta_b$, where $\rho_f$ is the liquid density) is of order $10^{-1}$ or lower, indicating that our simulations are in the laminar flow regime. For particle removal by water drops on hydrophobic surfaces, the interaction is a competition between the capillary force and the friction force~\citep{naga_how_2021}. To characterise the interplay between the friction force and the capillary force, we introduce the dimensionless number $\mu g_p/(\sigma R_p)$, where the numerator is the friction force that the particle experiences when sliding on a (dry) solid surface and the denominator is a measure of the capillary force on the particle. In the following, we varied this parameter between $0\leq\mu g_p/(\sigma R_p)\leq26$. 

\subsection{Push--pull versus enter--exit scenarios}

\begin{figure}
  \centerline{\includegraphics[width=0.9\linewidth]{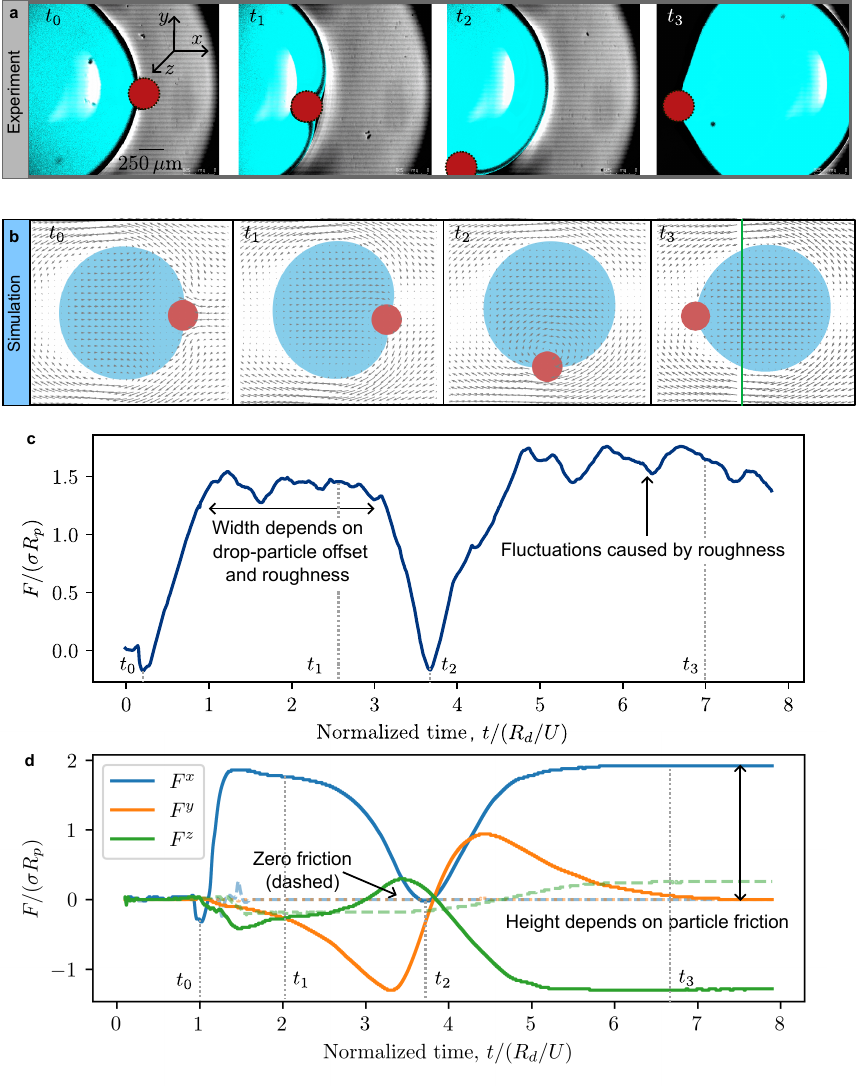}}
  \caption{Push--pull scenario. (a) Experiments showing the push-pull scenario when a water drop collides with a hydrophobic particle on a hydrophobic surface at 50\,$\mathrm{\mu m/s}$. The snapshots show the bottom-view of the collision in a horizontal plane through the centre of the particle. Reproduced from \citet{naga_how_2021} under a CC BY 3.0 licence. (b) Simulations of a drop colliding with a particle with low friction, as seen in a horizontal plane through the centre of the particle. The velocity field is overlaid on top, with the arrows corresponding to velocity in the drop's centre-of-mass frame. Green line in $t_3$ indicates where two images have been stitched together when the drop is at the periodic boundary. (c) Experimental force (along $x$) curve for the force on the particle. Data from \citet{naga_how_2021}. (d) Force that drop exerts on particle (\textit{i.e.} capillary + hydrodynamic forces), obtained with our LBM--DEM method. Different colours show force components in different directions. When the particle is frictionless, the forces along the $x$ and $y$ directions are zero (dashed lines) because there is no resistance to lateral motion. The drop still exerts a vertical force on the particle due to the asymmetric shape of the liquid meniscus around the particle. In Supplementary Figure 2, we split the forces into contributions due to the hydrodynamic force and the capillary force.}
\label{fig:PushPull}
\end{figure}

\begin{figure}
  \centerline{\includegraphics[width=0.9\linewidth]{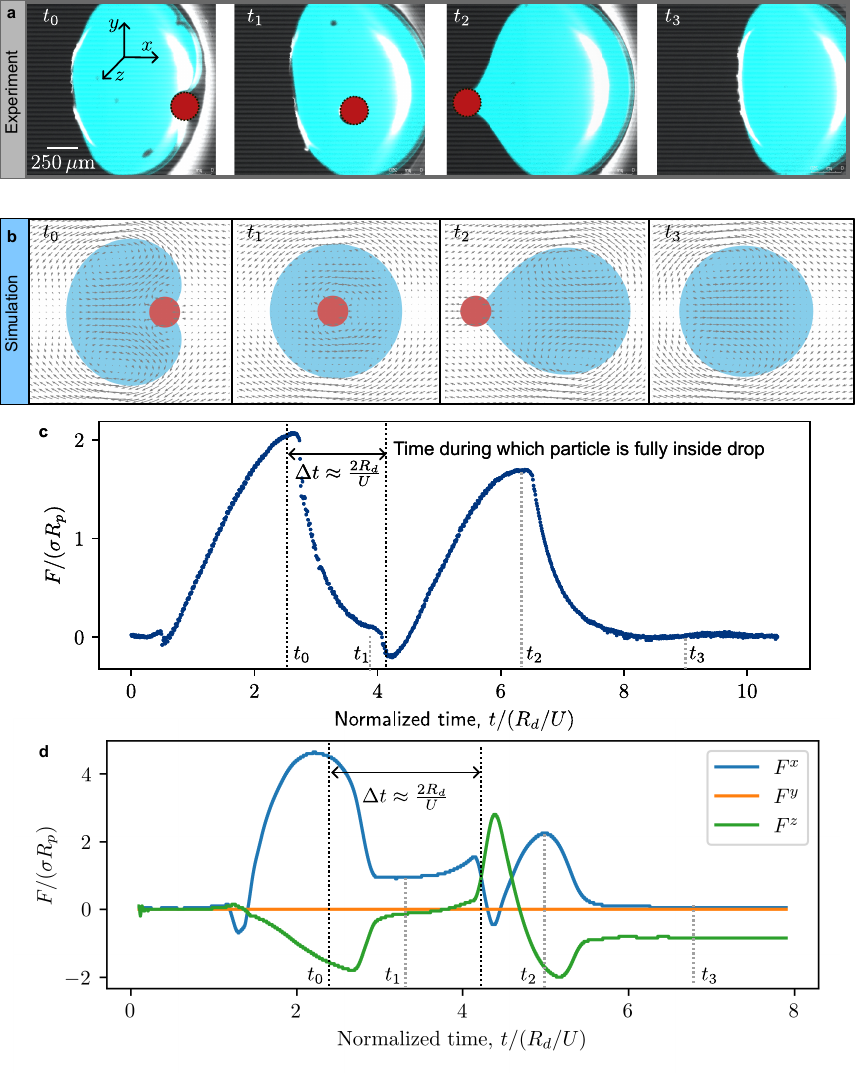}}
  \caption{Enter--exit scenario. (a) Experiments showing the enter--exit scenario when a water drop collides with a hydrophobic particle on a hydrophobic surface at 500\,$\mathrm{\mu m/s}$. The snapshots show the bottom-view of the collision in a horizontal plane through the centre of the particle. Reproduced from \citet{naga_how_2021} under a CC BY 3.0 licence. (b) Simulations of a drop colliding with a particle with high friction, as seen in a horizontal plane through the centre of the particle. The velocity field is overlaid on top, with the arrows corresponding to velocity in the drop's centre of mass frame. In both experiments and simulations, the particle enters and exits the drop. (c) Experimental force curve for the force (along $x$) acting on the particle during the collision. Data from \citet{naga_how_2021}. (d) Force that drop exerts on particle (\textit{i.e.} capillary + hydrodynamic forces), obtained with our numerical method. The force components in all three Cartesian directions can be obtained, as shown by the different colours. In both experiments and simulations, the force is maximum when the drop crosses the drop-air interface, highlighting that capillary forces dominate. Here $F_z$ remains negative after the particle exits because a small amount of liquid is left behind with the particle, leading to an attractive capillary force between the particle and the surface.}
\label{fig:EnterExit}
\end{figure}

To test whether our numerical model captures the key interactions necessary to study particle removal by drops, we simulated the collision between a drop and a particle on a surface and compared our results with previous experiments~\citep{naga_how_2021}. The experiments were performed with water drops and spherical hydrophobic glass particles ($R_p\approx 115\,\mathrm{\mu m}$) on a hydrophobic polydimethylsiloxane (PDMS) surface, as described in~\citet{naga_how_2021}. Briefly, the collision between the drop and the particle was imaged using an inverted laser scanning confocal microscope. The drop was fixed within the field of view of the microscope using a metallic cantilever while the surface moved at controlled speeds. With this setup, it is possible to image the collision between the drop and the particle over distances of several centimetres, which is much larger than the millimetric field of view of the microscope. The experimental imaging provided horizontal slices of the collision, as shown in figure\,\ref{fig:PushPull}a. Furthermore, the horizontal force $F_x$ acting on the drop and the particle along $x$ could also be measured directly from the deflection $\Delta x$ of the cantilever and using Hooke's law, $F_x=k_s\Delta x$, where $k_s$ is the spring constant of the cantilever. The contact angle between the drop and the surface was between $80^\circ$ and $120^\circ$, and the contact angle between the drop and the particle was between $75^\circ$ and $100^\circ$. The range spanned by these angles is due to contact angle hysteresis caused by roughness, chemical heterogeneities, and surface adaptation effects. For simplicity, our numerical model does not include static contact angle hysteresis. However, it still captures dynamic contact angle hysteresis.

\begin{figure*}
  \centerline{\includegraphics[width=1.0\linewidth]{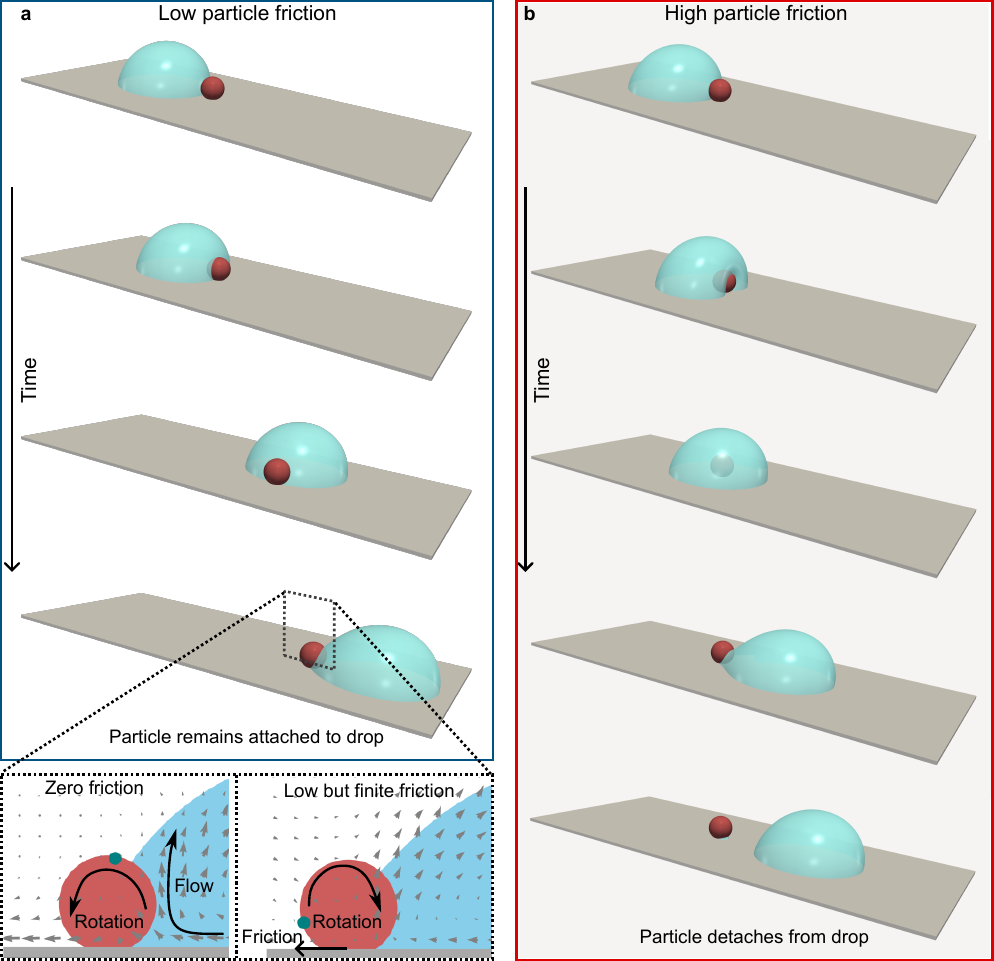}}
  \caption{Two types of collision scenarios between a drop and a particle on a horizontal surface. (a) Push--pull scenario. When the friction between the particle and the surface is small compared with the maximum capillary force that the drop can exert on the particle, the particle moves along the circumference of the drop and remains attached to the rear. In the example shown here, the particle moves clockwise around the drop because it is initially positioned slightly to the right-hand side of the drop. When there is no initial offset, the drop pushes the particle, and the latter remains in front of the drop for much longer. The insets show that the particle rolls when it is pulled by the drop. Frictionless particles roll anticlockwise, following the flow in the drop (arrows show fluid flow as viewed in the drop's centre of mass frame). Frictional particles roll in the opposite direction (clockwise) due to friction between the particle and the surface. (b) Enter--exit scenario. When the friction force exceeds the maximum capillary force, the particle penetrates the drop, travels across its base, and detaches from the drop at the rear side.}
\label{fig:3dsnapshots}
\end{figure*}

Experimentally, two scenarios were reported for the collision between a water drop and a hydrophobic particle. In the first scenario, which we call the push--pull scenario, the particle moved around the drop's footprint and remained attached to the rear drop-air interface (figure \ref{fig:PushPull}a). In the second scenario (enter--exit scenario), the particle entered and exited the drop (figure \ref{fig:EnterExit}a). Both the push--pull and enter--exit scenarios were imaged directly using laser scanning confocal microscopy~\citep{naga_how_2021}. However, due to the limited temporal resolution of this technique, the imaging was restricted to a single 2-D horizontal plane. By tuning the friction coefficients, we could successfully reproduce these two scenarios using our LBM--DEM method (figure \ref{fig:PushPull}b and figure \ref{fig:EnterExit}b). Our numerical method also allows us to visualise the interaction in three dimensions for the first time (figure\,\ref{fig:3dsnapshots}, Supplementary movies 1 and 2).

To simulate the two scenarios, we positioned a spherical particle on the substrate in front of the drop. The contact angle between the drop and the particle and between the drop and the surface were both set to $90^\circ$. We then applied a constant force to the drop to move it towards the particle. We chose the applied force, drop viscosity, and surface tension such that the capillary force between the drop and the particle dominates the hydrodynamic force, as is typically the case in experiments using water drops \citep{naga_how_2021}. This regime corresponds to when the capillary number of the drop is significantly less than unity. In these simulations, the capillary number was of the order of $\mathrm{Ca}\approx10^{-2}$. Our numerical method successfully reproduced the push-pull scenario in the limit where the friction coefficients $\mu$ and $\mu_r$ were both small, such that the capillary force between the drop and the particle exceeds the friction force between the particle and the surface. The enter-exit scenario was reproduced in the opposite limit when the friction force exceeds the capillary force.

In the push--pull scenario, the particle goes around the drop because any slight offset in the initial alignment between the particle and the drop relative to the direction of motion causes the particle to move sideways due to the convex shape of the front of the drop. Once at the rear of the drop, the particle remains attached to the drop-air interface. The particle always ends up in the same final configuration at the rear, regardless of the initial offset between the drop and the particle.

To compare the experiments and simulations quantitatively, we analysed the forces acting on the particle. In the experiments, the force acting on the particle alone was obtained by subtracting the total measured force (on drop $+$ particle) from the force acting on the drop alone before it collided with the particle. In the simulations, the force that the drop exerts on the particle can be obtained directly. While current experimental methods only provide the force acting along the direction of motion ($x$), our numerical method provides the force in all three Cartesian directions ($x$, $y$, $z$, as defined in figure~\ref{fig:PushPull}a, $t_0$).

The experimental and numerical force curves show good agreement (Figure\,\ref{fig:PushPull}c and d, respectively), indicating that the numerical model captures all the key physical interactions. The force curve along $x$ is characterised by two broad plateaus, corresponding to when the particle is in front and behind the drop. The first plateau is due to the force exerted by the drop as it pushes the particle, overcoming friction forces. The width of the first plateau depends on the amount of time that the particle spends in front of the drop before moving to the side. When there is an initial offset in the alignment of the drop and the particle (as is the case in figure\,\ref{fig:PushPull}b and \ref{fig:PushPull}d), the force in the $y$ direction is non-zero (orange curve in figure\,\ref{fig:PushPull}d). This non-zero force causes the particle to move around the drop perimeter.  When the particle is at the side (time $t_2$), the $x$ component of the force decreases to zero since at that instant, the drop does not push the particle in the forward direction. The second plateau in the horizontal force corresponds to when the particle is pulled at the rear of the drop. Interestingly, the drop also exerts a vertically downward force when pulling the particle (green curve in figure\,\ref{fig:PushPull}d).

Macroscopically, the push-pull scenario could be reproduced even without accounting for friction. When both $\mu$ and $\mu_r$ were set to zero, the scenario looked similar to the one shown in figure\,\ref{fig:PushPull}a,b. However, despite the macroscopic similarity, there are significant differences in the force curves (dashed lines in figure~\ref{fig:PushPull}d show force curves for a frictionless particle). Since no force is required to push or pull a frictionless particle, the forces along the $x$ and $y$ directions were always zero (dashed blue and orange lines). The absence of a horizontal force also meant that the drop did not deform as much when pushing or pulling a frictionless particle. However, a small vertical force was still present (dashed green line) due to the asymmetric shape of the water meniscus around the particle. Therefore, although ignoring friction leads to a similar macroscopic outcome, it is crucial to include friction to reproduce the forces on the particle accurately.

An important difference between the simulations and experiments is the presence of random roughness on the surface and on the particle in experiments. Roughness causes the particle to roll unevenly on the surface, which leads to fluctuations in the plateaus in the force curve when the drop pushes and pulls the particle, as seen in figure~\ref{fig:PushPull}c. Roughness also causes the particle to move around the drop perimeter even when the particle is initially positioned in perfect alignment with the drop. Random fluctuations in the particle motion due to roughness cause the particle to break the left--right symmetry and go around the drop. In contrast, in the simulations, the particle can be pushed in front of the drop until the left--right symmetry is broken by numerical noise~\citep{connington_review_2012} in the particle position. For the comparison with experiments in figure~\ref{fig:PushPull}d, we introduced a small misalignment by positioning the particle slightly to the right of the drop, to favour clockwise motion as it moves around the drop perimeter.

For the enter--exit scenario, the force curve (for force along $x$) is characterised by two peaks, shown in figure~\ref{fig:EnterExit}(c,d). The maxima of these peaks correspond to when the particle crosses the front and rear drop-air interface. The time interval between the maximum of the first peak and the beginning of the second peak corresponds to the time taken for the particle to travel across the drop footprint, $\Delta t\approx2L/U$, where $L$ is the length of the drop along the direction of motion and $U$ is the relative velocity between the drop and the particle.

When the particle moved across the footprint of the drop, the force reduced significantly compared with when the particle crossed the interface. This reduction in force is because when the particle is fully submerged, the capillary force is zero, and the particle only experiences hydrodynamic force due to viscous drag. For slow speeds, as was the case in the experiments where the speed was $\mathcal{O}(10^{-4}\,\mathrm{m/s})$, the viscous drag force is small (estimated to be $\mathcal{O}(10^{-9}\,\mathrm{N})$~\citep{naga_how_2021}) compared with the capillary force when the particle is at the liquid-air interface. Consequently, the force reduced to $\approx 0$ between the two peaks in the experimental force curve. In the simulations, the reduction in force was less significant because we used a higher drop speed $\mathcal{O}(1\,\mathrm{m/s})$ to keep computational costs tractable. Due to the higher drop speed in the simulations, the force along $x$ does not decrease completely to zero when the particle moves inside the drop. Note that despite the four orders of magnitude difference in the absolute drop speed, the Reynolds number of the particle only differs by one order of magnitude in the experiments ($\mathrm{Re}\approx 10^{-2}$) and simulations ($\mathrm{Re}\approx10^{-1}$) because the difference in speed is compensated by having a kinematic drop viscosity $\eta_b/\rho_f$ that is around three orders of magnitude higher in simulations. Crucially, in both experiments and simulations, the same hierarchy of forces applies, with the capillary force being typically larger than the viscous force ($\mathrm{Ca}\ll1$), which is itself larger than the inertial force ($\mathrm{Re}<1$).

Our simulations further indicate that the drop exerted a vertical force on the particle when the liquid-air interface moved across the particle. Figure~\ref{fig:EnterExit}d shows that the vertical force was negative throughout the interaction, except when the particle first reached the rear drop-air interface due to dewetting of the liquid on the particle. The vertical force also remained negative even after the particle exited the drop (time $t_3$ in figure~\ref{fig:EnterExit}d) due to the presence of a small amount of liquid being left under the particle after the detachment. This remnant liquid leads to capillary adhesion between the particle and the surface.

\begin{figure*}
  \centerline{\includegraphics[width=1.0\linewidth]{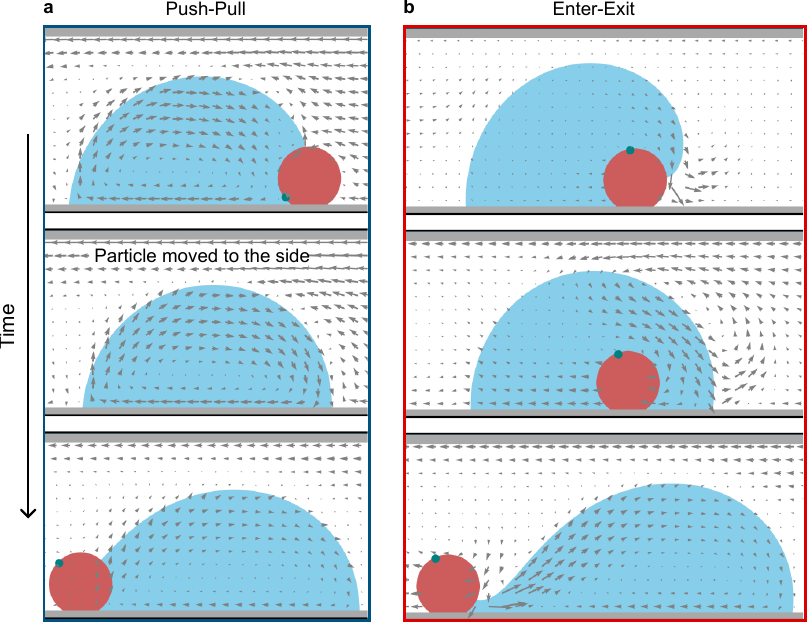}}
  \caption{Velocity profile in the drop and surrounding phase for the push-pull (a) and enter--exit (b) scenarios. These velocity profiles are taken in a vertical slice close to the centre of mass of the drop. The velocities shown are in the centre of mass frame of the drop. In (a), there is a small but finite friction force. In (b), there is a large friction force. A satellite drop becomes entrained by the particle due to the breakup of the capillary bridge when it exits the drop.}
\label{fig:vertical slice velocity}
\end{figure*}

In addition to enabling 3-D visualisation and providing force measurement in all directions, another benefit of the numerical method is that it allows us to obtain flow profiles. In both the push-pull and enter-exit scenarios, the flow in the drop followed a clockwise rolling motion when viewed in the centre of mass frame of the drop, as shown in figure \ref{fig:vertical slice velocity}a,b. Supplementary movies 3, 4 and 5, 6 show flow profiles in vertical and horizontal slices across the centre of the particle, respectively. Interestingly, the flow direction in the drop is opposite to the direction in which friction causes the particle to rotate. When the drop pulls the particle, flow promotes anticlockwise rotation of the particle, whereas friction between the particle and the surface promotes clockwise rotation, as shown in the inset at the bottom of figure \ref{fig:3dsnapshots}a. Consequently, frictional particles roll clockwise, driven by the friction force between the particle and the surface, whereas frictionless particles roll anticlockwise, driven by the fluid flow in the drop. This effect reinforces the conclusion that even when friction is small, it cannot be ignored since it plays a crucial role in determining the microscopic motion of the particle.

\begin{figure}[h!]
  \centerline{\includegraphics{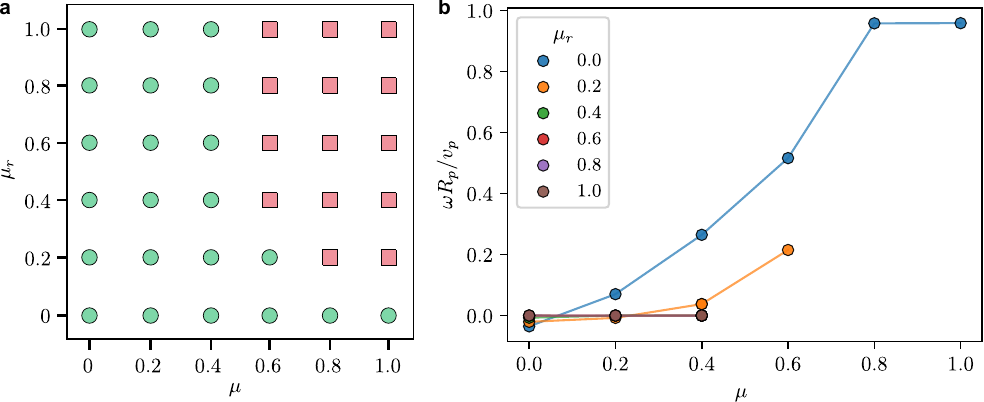}}
  \caption{(a) Outcomes of the collision between a drop and a particle as $\mu$ and $\mu_r$ are varied independently. Here, $\mathrm{Bo}=1.25$ and $0\leq \mu g_p/(\sigma R_p)\leq 5.2$. The green circles correspond to the case where the particle remains attached to the drop, as shown in figure\,\ref{fig:3dsnapshots}a. The red squares correspond to when the particle detaches from the drop. (b) Variation of the rotational velocity of the particle with the friction coefficient. The rotational velocity is normalised by the centre of mass velocity of the particle. A ratio of 1 indicates pure rolling, whereas a ratio of 0 indicates pure sliding. Even when the rolling friction is set to zero ($\mu_r=0$), pure rolling is not the preferred mode of motion, and there remains some relative sliding between the particle and the surface up to $\mu=0.8$. When $\mu_r=\mu$, the particle slides rather than rolls. The rotational velocity can be negative when $\mu=0$, indicating that the particle rotates anticlockwise. This anticlockwise rotation is driven by the anticlockwise flow of liquid inside the drop.}
\label{fig:varymu}
\end{figure}

\subsection{Transition from push--pull to enter--exit scenario}

We showed that the friction force between the particle and the surface determines whether the collision follows the push-pull or enter-exit scenarios. Yet, there are several other factors that may influence these scenarios. First, there are two types of friction that enter this problem: sliding friction and rolling friction. How do these two types of friction influence the collision scenario? Furthermore, even when the rolling and sliding friction coefficients are fixed, the collision scenario may depend on the drop size or the tilt angle of the surface. In the following, we investigate the effects of varying the coefficients of sliding and rolling friction, the drop size, and the force applied to the drop.

To investigate the role of sliding and rolling friction, we varied the coefficients of sliding and rolling friction ($\mu$ and $\mu_r$, respectively) independently between 0 and 1 and mapped out whether the particle remained attached to the drop (figure~\ref{fig:varymu}a). In this set of simulations, we kept everything else constant, including the contact angles, particle weight, and force applied to the drop. It turns out that the particle could always be captured when at least one of the friction coefficients was zero (figure~\ref{fig:varymu}a). However, this was no longer the case when the friction coefficients were finite.

As shown in figure\,\ref{fig:varymu}a, the outcome depends on both the sliding and rolling friction coefficients. However, for successful particle capture, it is more important for the sliding friction to be small because the distribution of outcomes is not symmetric about the line $\mu_r=\mu$. A higher number of successful capture cases lies above the $\mu_r=\mu$ line, indicating that having a high $\mu$ is more detrimental to capture than having a high $\mu_r$. For example, when $\mu=1.0$, the particle is never captured for any $\mu_r>0$. However, when $\mu_r=1.0$, the particle can still be captured as long as $\mu<0.6$. Therefore, although both sliding and rolling friction are important to predict whether a particle can be captured, the sliding friction plays a larger role in determining the outcome.

To understand when the particle rolls or slides on the surface, we analysed the angular velocity of the particle for the cases when it remained attached at the rear of the drop (figure\,\ref{fig:varymu}b). When both the sliding and rolling friction were zero, the angular velocity of the particle was slightly negative, indicating an anticlockwise rotation in the direction of the flow in the drop, as discussed earlier in figure\,\ref{fig:3dsnapshots}. However, as soon as the sliding friction was non-zero, the angular velocity switched sign, indicating a clockwise rotation driven by sliding friction between the particle and the surface. When the sliding and rolling friction coefficients were equal, sliding was the preferred mode of motion. In this case, the preference for sliding can be understood to be because there are additional sources of rolling resistance that enhance the rolling friction beyond the sliding friction, including the rolling resistance experienced by the particle when it rotates against the opposing liquid flow inside the drop and due to the resistive capillary torque~\citep{naga_capillary_2021} associated with the particle rotating at a liquid-air interface. Increasing $\mu$ beyond $\mu_r$ reduced the relative amount of sliding between the particle and the surface, until the particle started to display a pure rolling motion. Pure rolling corresponds to when the rotational velocity and translational velocity become equal, $\omega_p R_p=v_p$, such that there is no relative sliding between the particle and the surface at the point of contact. For a constant sliding friction coefficient, the amount of rolling relative to sliding decreases with an increase in the rolling friction coefficient (the orange compared with the blue curve in figure\,\ref{fig:varymu}b).

Experimentally, it was observed that a spherical hydrophobic particle rolls on a hydrophobic surface without any noticeable sliding when it is pulled by a water drop~\citep{naga_how_2021}. In experiments, rolling was deduced by tracking the position of defects on the particle. Based on the findings shown in figure\,\ref{fig:varymu}b, our numerical results reveal that, in general, the amount of rolling relative to sliding depends on the relative size of the coefficients of rolling and sliding friction. However, since in experiments, the rolling friction coefficient for spherical particles is typically much smaller than the sliding friction coefficient, spherical particles are much more likely to roll.

\begin{figure}
  \centerline{\includegraphics[width=0.9\linewidth]{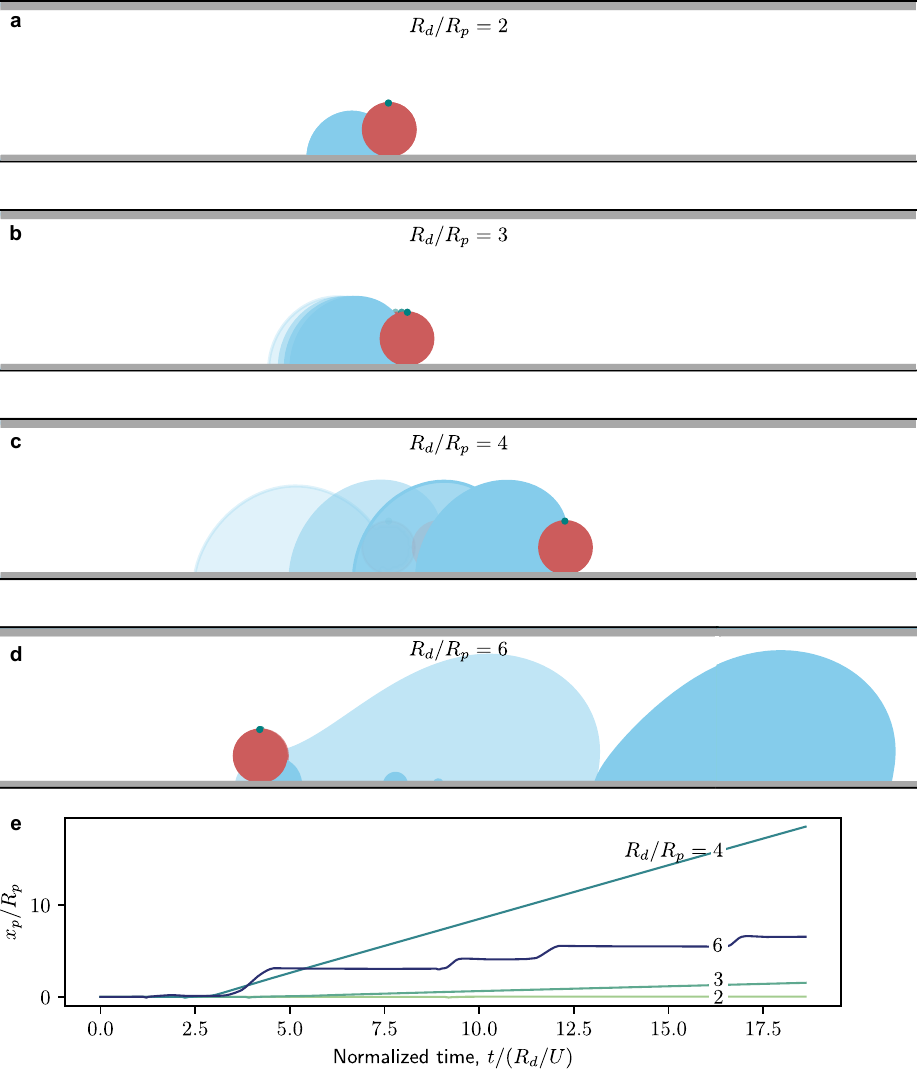}}
  \caption{(a-d) Time-lapse showing the effect of increasing the drop radius $R_d$ relative to the particle radius, $R_p=100\,\mathrm{\mu m}$. In all cases, $\mu g_p/(\sigma R_p)=5.23$, $\mu=\mu_r$, and the drop Bond number varies between $0.16 \leq \mathrm{Bo}\leq 1.44$. The force applied per unit drop volume is the same in all cases, therefore changes in $\mathrm{Bo}$ are due to changes in drop radius. Translucent images indicate earlier times. The particle was positioned in perfect alignment with the drop. Therefore, within the simulated timescale, it remained in front of the drop and did not go around. In (d), the particle enters and exits the drop. In this case, satellite droplets are left behind when the particle detaches from the main drop. (e) Position of the particle as a function of time for the cases shown in (a-d). Time is normalised using the initial drop radius and the characteristic drop speed corresponding to the enter--exit case when $R_d/R_p=6$. When $R_d/R_p=2$, the force exerted by the drop is insufficient to push the particle. For the largest drop ($R_d/R_p=6$), the particle only moves when it crosses the front and rear drop--air interface. There are four steps in the curve because the drop collides with the particle a second time after crossing the periodic boundary of the simulation domain.}
\label{fig:vary_drop_vol}
\end{figure}

\subsection{Effects of drop size and speed}
\label{vary size and speed}

In practical scenarios and cleaning applications, we are likely to encounter drops of different sizes moving at different speeds. Thus, it is important to understand how the drop size and speed affect its ability to capture a particle. To understand how the removal mechanism is influenced by the drop size, we varied the drop size while keeping the particle size constant (figure~\ref{fig:vary_drop_vol}). In all cases, we applied the same force per unit volume to move the drop. Therefore, the total force applied to the drop was proportional to its volume. This procedure mimics drops of different volumes sliding down a vertical surface under the action of gravity.

The drop size influences its ability to capture as well as transport the particle (figure~\ref{fig:vary_drop_vol}). When the drop was too small ($R_d/R_p=2$, figure~\ref{fig:vary_drop_vol}a), it was unable to displace the particle because the total force exerted by the drop was insufficient to overcome the friction force on the particle. As the drop volume increased ($R_d/R_p=3,4$, figure~\ref{fig:vary_drop_vol}b,c), the drop successfully displaced the particle by pushing it forward. Here, within the simulated timescale, the particle remained in front of the drop rather than going around (as shown previously in figure~\ref{fig:PushPull}) because it was positioned in perfect alignment with the drop. Therefore, a larger drop improves removal efficiency because it moves faster and can transport the particle over a larger distance within the same time interval.

However, drops that are too large are no longer efficient at transporting particles. As shown in figure\,\ref{fig:vary_drop_vol}d, when $R_d/R_p=6$, the particle entered and exited the drop. When the particle enters and exits the drop, it can no longer be displaced efficiently because it only moves during the brief periods when it interacts with the liquid--air interface (figure~\ref{fig:vary_drop_vol}e). This transition to the enter-exit scenario can be understood because larger drops move faster, causing the dynamic advancing and receding contact angles at the front and rear of the drop to change, as seen by the greater drop deformation in figure\,\ref{fig:vary_drop_vol}d compared with figure\,\ref{fig:vary_drop_vol}c. Changes in the contact angles influence the capillary force between the drop and the particle~\citep{naga_thesis_2021} and can therefore influence the outcome of the collision. Therefore, particles can only be efficiently transported by drops that are neither too large nor too small, highlighting that there is an optimal drop size. In general, the optimal drop size will depend on the friction coefficients, contact angles, and the surface tension of the liquid. All these quantities must be taken into account when optimising the drop size to enable efficient particle capture and removal on surfaces.

\begin{figure}[h!]
  \centerline{\includegraphics[width=0.9\linewidth]{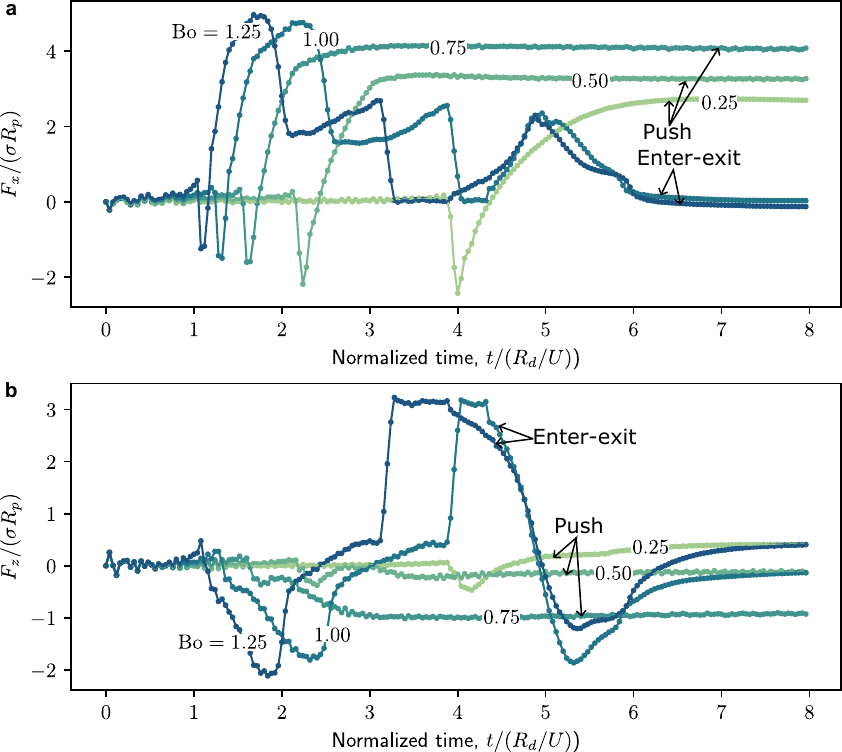}}
  \caption{Varying the force per unit volume applied to the drop. For all cases here, the drop volume is constant, $\mu g_p/(\sigma R_p)=5.23$, $\mu=\mu_r$, and the particle is perfectly aligned with the drop. (a) Force acting on the particle along $x$ for different drop speeds. For the lowest three $\mathrm{Bo}$, the drop pushes the particle, and therefore the force curves have a single plateau, corresponding to the first plateau in the push--pull scenario. For the highest two $\mathrm{Bo}$, the particle enters and exits the drop. The two peaks in the force curves correspond to when the particle crosses the front and rear liquid-air interface, respectively. (b) Vertical force (along $z$) acting on the particle corresponding to the five cases in (a). The drop exerts a downward force on the particle when the particle crosses the front and rear liquid-air interface. In both (a) and (b), the time is normalised by $R_d/U$, where $U$ is the approximate drop speed for the two enter--exit scenarios shown here.}
\label{fig:vary_drop_force}
\end{figure}

Besides optimising the drop size, it is also important to consider how the tilt angle of the surface affects the collision scenario for a fixed drop volume. Varying the tilt angle changes the force acting on the drop parallel to the surface. Therefore, to understand the effect of varying the tilt angle, we varied the force applied to the drop parallel to the surface. In figure\,\ref{fig:vary_drop_force}, we compare five cases corresponding to different forces applied to the drop, where the applied force is indicated by the Bond number. For these simulations, the drop and particle were positioned in perfect alignment. Therefore, when the particle did not enter the drop, it remained in front of the drop within the simulated timescale. As the applied force was increased, the collision scenario transitioned from the particle being pushed by the drop to it entering and exiting the drop. For the lowest three applied forces, the particle was pushed by the drop, whereas for the highest two forces, it entered and exited the drop. Comparing the force curves for the first three and last two cases separately allows us to understand how changes in the force applied to the drop influence the force on the particle for the two different scenarios.

When the particle remained in front of the drop, a higher force was required to push the particle as the force applied to the drop increased (figure\,\ref{fig:vary_drop_force}a). The increase in the horizontal force required to push the particle was correlated with an increase in the downward vertical force acting on the particle (figure\,\ref{fig:vary_drop_force}b). Since the friction force experienced by the particle is proportional to the normal reaction force between the particle and the surface, the increase in the downward force causes the friction to increase. As a consequence, when the drop exerts a downward force on the particle (\textit{e.g.} due to a change in the contact angle or in the flow profile), it also needs to exert a larger horizontal force to push the particle.

As the applied force increased, the horizontal force required to push the particle increased until the drop could no longer push the particle. For the highest two applied forces ($\mathrm{Bo}=1.00$ and $1.25$), the particle entered and exited the drop. The horizontal force on the particle as it entered the drop was almost the same for these two cases (figure\,\ref{fig:vary_drop_force}a, first peak of the top two curves). Once the particle had fully crossed the front drop-air interface, it experienced hydrodynamic drag before reaching the rear drop--air interface. When the particle reached the rear drop--air interface, the horizontal force initially decreased as it snapped into contact with the interface before increasing again as it detached from the interface. When the particle exited the drop, the horizontal force on the particle was almost the same for $\mathrm{Bo}=1.00$ and $1.25$. Overall, these results indicate that the tilt angle of the surface can affect particle capture because the capillary force exerted by the drop–air interface is limited by a threshold value determined by the surface tension and contact angles~\citep{naga_how_2021}, and cannot increase indefinitely to overcome friction.

\section{Conclusions}

In this paper, we introduced a 3-D numerical method coupling a binary fluid LBM and a DEM to model interfacial flows interacting with frictional solid particles. Our method explicitly accounts for the hydrodynamic force in the bulk, the capillary force at liquid-fluid interfaces, and normal, sliding, and rolling contact forces between particles and solid surfaces.

We benchmarked each force independently. We benchmarked the hydrodynamic force by investigating the trajectory of a spherical particle in a Poiseuille flow. For the capillary force, we accurately reproduced the force required to detach a particle from a liquid-fluid interface for a wide range of contact angles. We validated the normal contact force and frictional forces by investigating the motion of a particle bouncing on a surface and sliding/rolling on a surface in the absence of fluid. All these benchmarks show good agreement with experiments and/or theoretical predictions.

To further demonstrate the capabilities of our LBM--DEM method, we applied the method to investigate how a drop captures and removes a particle on a solid surface. The method successfully reproduced experimentally observed removal scenarios and demonstrates that friction plays a crucial role in determining both the particle's rotational dynamics and the force between the drop and the particle. In particular, neglecting rolling and sliding friction leads to different rotational dynamics and force curves. Interestingly, our results indicate that for transporting particles efficiently, drops must be neither too small nor too large. A drop that is too small is inefficient because it moves slowly, whereas a drop that is too large is also inefficient because it moves across the particle, leaving it behind.

Our method enables independent variation of rolling and sliding friction coefficients, contact angles, surface tension, and viscosity, allowing systematic exploration of the parameter space governing particle removal. A key goal for future work is to derive a criterion to predict whether a drop can remove a particle on a surface across a wide range of particle and surface material properties. Our method opens up several areas for future investigation, including to understand rain-induced soil erosion~\citep{vaezi_contribution_2017}, self-assembly of colloidal particles~\citep{zargartalebi_self-assembly_2022}, how particle (virus)-laden respiratory drops interact with surfaces~\citep{seyfert_stability_2022}, or how raindrops transport microplastics in the environment~\citep{brahney_plastic_2020}. There are also important directions to extend the model, such as generalising the method beyond spherical particles, including textured surfaces (\textit{e.g.} pillars) to account for roughness, and incorporating deformable particles.

\section*{Acknowledgement}
We are grateful to Chris Ness for valuable discussions on modelling friction forces between particles, and to Yang Zhang and Sheng Li for thoroughly proofreading the paper. We also thank Mike Cates for feedback on the method and Glen McHale for regular discussions.

\section*{Funding}
This work was funded by an EPSRC National Fellowship in Fluid Dynamics (NFFDy) with grant number EP/X028410/2 (A.N.), an EPSRC Early Career Fellowship with grant number EP/V034154/2 (H.K., J.Y.), and a Leverhulme Trust Research Project Grant RPG-2022-
140 (X.Z., H.K.).

\section*{Declaration of interests}
The authors report no conflict of interest.

\section*{Data Availability Statement}
The data that support the findings of this study are available from the corresponding authors upon request.

\section*{Author ORCIDs}
A. Naga, https://orcid.org/0000-0001-7158-622X; H. Kusumaatmaja, https://orcid.org/0000-0002-3392-9479.

\section*{Author contributions}
A.N. and H.K. conceptualised the method to combine LBM and DEM. A.N. led the research project and coupled the LBM and DEM algorithms. X.Z. wrote the original model to compute the capillary force on particles. J.Y. carried out the benchmarks for the hydrodynamic force. A.N. wrote the manuscript, with contributions from all authors. All authors discussed the results and reviewed the manuscript.

\bibliographystyle{rsc}
\bibliography{references}

\end{document}